\documentclass[11pt,a4paper]{article}
\pdfoutput=1
\usepackage{jheppub}
\usepackage{slashed}
\usepackage[makeroom]{cancel}

\makeatletter
\def\@fpheader{\relax}
\makeatother

\usepackage{ulem}
\usepackage{tensor}
\usepackage[czech,english]{babel}
\usepackage{graphicx}
\usepackage{amsmath,amsfonts,amssymb}
\usepackage{url}
\usepackage{mathabx}
\DeclareMathOperator{\MyProd}{\scalebox{1.4}{$\mathrm{I\kern-0.2ex I}$}}

\usepackage{leftidx}
\usepackage{xcolor}
\numberwithin{equation}{section}
\usepackage{adjustbox}
\usepackage{appendix}
\usepackage{tikz}
\tikzstyle{process} = [rectangle, minimum width=3cm, minimum height=1cm, text centered, draw=black, fill=orange!30]
\tikzstyle{arrow} = [thick,->,>=stealth]

\preprint{LCTP-20-23}

\title{Universal Entropy and Hawking Radiation of Near-Extremal AdS$_4$ Black Holes}

\author[a]{Marina David}

\emailAdd{mmdavid@umich.edu}

\author[a]{Jun Nian}

\emailAdd{nian@umich.edu}

\affiliation[a]{Leinweber Center for Theoretical Physics, University of Michigan, Ann Arbor, MI 48109, U.S.A.}

\abstract{We compute the Bekenstein-Hawking entropy of near-extremal asymptotically AdS$_4$ electrically charged rotating black holes using three different methods: (i) from the gravity solution, (ii) from the near-horizon Kerr/CFT correspondence and (iii) from the boundary conformal field theory. The results from these three different approaches match exactly, giving us a unique and universal expression for the entropy and the microstate counting of near-extremal AdS black holes via the AdS/CFT correspondence. In the second method, we extend the Kerr/CFT correspondence to the near-extremal case to compute the left and right central charges. We also use hidden conformal symmetry of the near-horizon geometry to compute the Frolov-Thorne temperatures. From the results of the near-extremal AdS$_4$ black hole entropy, we provide a microscopic foundation for Hawking radiation.
}


\arxivnumber{}

\newcommand{\bea}{\begin{eqnarray}}
\newcommand{\eea}{\end{eqnarray}}

\newcommand{\be}{\begin{equation}}
\newcommand{\ee}{\end{equation}}

\begin{document}

\maketitle

\section{Introduction}\label{sec:Introduction}

The groundbreaking work of Strominger and Vafa \cite{Strominger:1996sh} provided the first microscopic explanation for the entropy of certain asymptotically flat black holes, and inspired a considerable amount of work in regards to understanding the microstates of asymptotically flat black holes. In the last few years, this has been expanded on for asymptotically AdS black holes using the AdS/CFT correspondence \cite{Maldacena:1997re, Witten:1998qj}. In particular, for a special class of supersymmetric extremal AdS black holes, which we call BPS black holes, the Bekenstein-Hawking entropies can be reproduced from the corresponding boundary conformal field theories (CFT) \cite{Benini:2015eyy, Cabo-Bizet:2018ehj, Choi:2018hmj, Benini:2018ywd, Choi:2019miv, Kantor:2019lfo, Nahmgoong:2019hko, Choi:2019zpz, Nian:2019pxj, Cabo-Bizet:2019eaf, Bobev:2019zmz, Benini:2019dyp, Hosseini:2019lkt, Choi:2019dfu, Crichigno:2020ouj}.

\begin{figure}
\begin{center}
  \includegraphics[width=13cm]{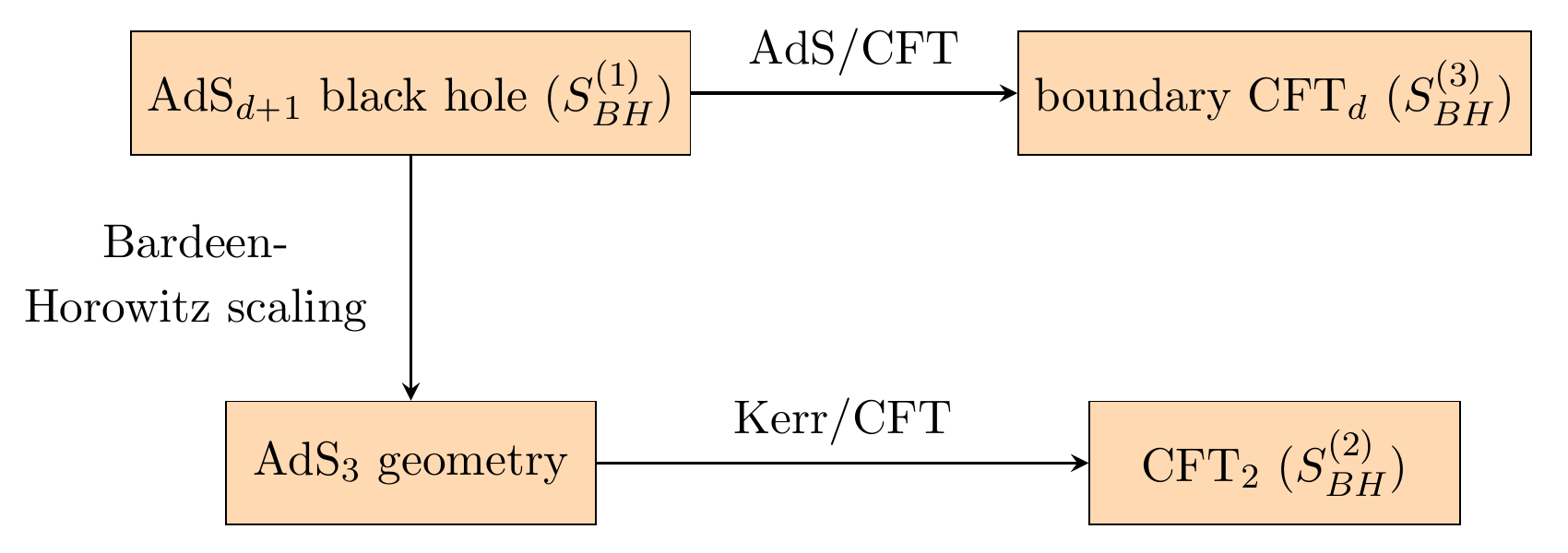}
  \caption{The asymptotically AdS black hole entropy can be computed in three different ways ($S_{BH}^{(i)}$), and have found to give one universal result for the entropy. This is valid for both BPS black holes and near-extremal black holes.} \label{fig:M1}
\end{center}
\end{figure}

Combining with the near-horizon Kerr/CFT correspondence \cite{Guica:2008mu, Lu:2008jk, Chow:2008dp}, we now have a unifying picture as shown in Fig.~\ref{fig:M1}. More precisely, for BPS black holes, their entropies can be computed in various ways, including (i) the original Bekenstein-Hawking formula on the gravity side, (ii) the near-horizon Kerr/CFT correspondence and (iii) the microstate counting from the boundary CFT via AdS/CFT correspondence. In addition, the useful limit called the gravitational Cardy limit can simplify the geometry near the horizon, by producing an AdS$_3$ subgeometry, which has been explicitly verified for AdS$_{4,5,6,7}$ BPS black holes in \cite{David:2020ems}.

Most of the previous work has assumed supersymmetry, and one natural question that may arise is whether or not the entropy matching is successful when the BPS bound - the intersection of supersymmetry and extremality is relaxed. In order to investigate this question, we study near-extremality in this paper. There has already been progress made in this direction, such as in \cite{Larsen:2019oll}, where AdS$_5$ black holes have been discussed from gravity, including near-extremal ($T_H > 0$) and extremal near-BPS ($T_H = 0$) configurations. For the latter case, the boundary field theory was also explored. Later, \cite{Nian:2020qsk} resolved some technical problems and successfully derived the heat capacity for near-extremal AdS$_5$ black holes from the boundary $\mathcal{N} = 4$ supersymmetric Yang-Mills theory (SYM). In addition, \cite{Nian:2020qsk} used the Bardeen-Horowitz near-horizon rescaling \cite{Bardeen:1999px} and the Kerr/CFT correspondence \cite{Guica:2008mu, Lu:2008jk, Chow:2008dp} to reproduce the near-extremal AdS$_5$ black hole entropy from the near-horizon CFT$_2$. Consequently, it provided a new microscopic formalism of Hawking radiation, similar to the D1-D5 CFT interpretation for asymptotically flat black holes studied by Callan and Maldacena \cite{Callan:1996dv}. In this work, we extend the analyses of \cite{David:2020ems} from the BPS case to the near-extremal case. However, unlike \cite{David:2020ems}, we do not assume the Cardy limit.

More explicitly, we compute the near-extremal electrically charged rotating AdS$_4$ black hole entropy using three different approaches:
\begin{enumerate}
\item expansion of the non-extremal AdS$_4$ black hole solution around the BPS solution,

\item near-extremal Kerr-Newman-AdS/CFT correspondence from the near-horizon CFT$_2$,

\item microstate counting via AdS/CFT correspondence from the boundary 3d superconformal ABJM theory at small temperature.
\end{enumerate}
We find that the entropy computations from the three different approaches lead to one universal result for the entropy, which can be formally written as
\be
  S_{BH} = S_* + \delta S = S_* + \left(\frac{C}{T_H}\right)_*\, T_H\, ,
\ee
where $S_*$ denotes the electrically charged rotating AdS$_4$ black hole entropy in the BPS limit, while $(C/T_H)_*$ stands for the heat capacity in the BPS limit. Therefore, our results use a concrete example to support the existence of a near-horizon CFT$_2$, which accounts for the black hole entropy, and show that the Kerr/CFT correspondence \cite{Guica:2008mu, Lu:2008jk, Chow:2008dp}, which was originally applied only to extremal black holes, can also be extended to the near-extremal black holes. In addition, the near-extremal AdS$_5$ black hole discussed in \cite{Nian:2020qsk} and the near-extremal AdS$_4$ black hole considered in this paper support that the unifying picture shown in Fig.~\ref{fig:M1} is valid for both BPS black holes and near-extremal black holes. As a byproduct, we clean up some technical issues in previous literature. This includes the computation of the right central charge of the near-horizon CFT$_2$ and that $c_L = c_R$ similar to \cite{Matsuo:2010ut}. In addition, we can unambiguously derive the right Frolov-Thorne temperature $T_R$ and consequently the black hole entropy from the right sector without any cutoffs, as in \cite{Chen:2010bh, Matsuo:2010ut}.

Several recent works based on 2d JT gravity coupled to a 2d bath CFT \cite{Penington:2019npb, Almheiri:2019hni, Almheiri:2019psf, Almheiri:2019qdq} have provided a holographic description of Hawking radiation and black hole evaporation, which can potentially be used to address the long-standing problem of black hole information paradox. This paper together with \cite{Nian:2020qsk} give an alternative approach of studying black hole evaporation and information paradox via AdS/CFT correspondence. In contrast to \cite{Penington:2019npb, Almheiri:2019hni, Almheiri:2019psf, Almheiri:2019qdq}, which are intrinsically 2d models, both \cite{Nian:2020qsk} and our work include both higher-dimensional CFTs on the boundary and a near-horizon 2d CFT, which may be more robust.

This paper is organized as follows. In Sec.~\ref{sec:AdS4} we carefully discuss how to achieve near-extremal black holes by perturbing the extremal condition for the supersymmetric black hole solutions, and then compute the near-extremal electrically charged rotating AdS$_4$ black hole entropy in three different ways. In Sec.~\ref{sec:HawkingRadiation} we connect the near-extremal AdS$_4$ black hole entropy obtained in the previous section with Hawking radiation, and provide a microscopic formalism similar to \cite{Callan:1996dv} for asymptotically flat black holes and \cite{Nian:2020qsk} for AdS$_5$ black holes. Some discussions for the future research directions are presented in Sec.~\ref{sec:Discussion}.

While this work was being completed, we became aware of the related work \cite{Larsen:2020lhg}, which has some overlap with Subsection~\ref{sec:SBH from Gravity} in this paper.

\section{Near-Extremal AdS$_4$ Black Hole Entropy}\label{sec:AdS4}

\subsection{AdS$_4$ Black Hole Solution}

In this subsection, we review the asymptotically AdS$_4$ black holes of interest along with some of their properties. The non-extremal asymptotically AdS$_4$ electrically charged rotating black hole solution with a gauge group $U(1) \times U(1)$ was constructed in \cite{Chong:2004na, Cvetic:2005zi} as a solution to 4d $\mathcal{N}=4$ gauged supergravity. This is a special case, where there are two electric charges that are pairwise equal ($Q_1 = Q_3$, $Q_2 = Q_4$) along with one angular momentum $J$. The solution is characterized by four parameters $(a, m, \delta_1, \delta_2)$. The metric, the scalar fields and the gauge fields are given by
\be
  ds^2 = - \frac{\Delta_r}{W} \left(dt - \frac{a\, \textrm{sin}^2 \theta}{\Xi} d\phi \right)^2 + W \left(\frac{dr^2}{\Delta_r} + \frac{d\theta^2}{\Delta_\theta} \right) + \frac{\Delta_\theta\, \textrm{sin}^2 \theta}{W} \left(a\, dt - \frac{r_1 r_2 + a^2}{\Xi} d\phi \right)^2\, ,\label{eq:AdS4Metric1}
\ee
\begin{align}
\begin{split}
  e^{\varphi_1} & = \frac{r_1^2 + a^2\, \textrm{cos}^2 \theta}{W}\, ,\qquad \chi_1 = \frac{a (r_2 - r_1)\, \textrm{cos}\, \theta}{r_1^2 + a^2\, \textrm{cos}^2 \theta}\, ,\\
  A_1 & = \frac{2 \sqrt{2} m\, \textrm{sinh} (\delta_1)\, \textrm{cosh} (\delta_1)\, r_2}{W} \left(dt - \frac{a\, \textrm{sin}^2 \theta}{\Xi} d\phi \right) + \alpha_{41}\, dt\, ,\\
  A_2 & = \frac{2 \sqrt{2} m\, \textrm{sinh} (\delta_2)\, \textrm{cosh} (\delta_2)\, r_1}{W} \left(dt - \frac{a\, \textrm{sin}^2 \theta}{\Xi} d\phi \right) + \alpha_{42}\, dt\, ,
\end{split}
\end{align}
where
\begin{align}
\begin{split}\label{eq:AdS4MetricFactors}
  r_i & \equiv r + 2 m\, \textrm{sinh}^2 (\delta_i)\, ,\quad (i = 1, 2)\\
  \Delta_r & \equiv r^2 + a^2 - 2 m r + g^2 r_1 r_2 (r_1 r_2 + a^2)\, ,\\
  \Delta_\theta & \equiv 1 - g^2 a^2\, \textrm{cos}^2 \theta\, ,\\
  W & \equiv r_1 r_2 + a^2\, \textrm{cos}^2 \theta\, ,\\
  \Xi & \equiv 1 - a^2 g^2\, ,
\end{split}
\end{align}
and $g \equiv \ell_4^{-1}$ is the inverse of the AdS$_4$ radius. We have added two pure gauge terms to the two gauge fields $A_{1, 2}$ with constants $\alpha_{41}$ and $\alpha_{42}$. The metric \eqref{eq:AdS4Metric1} can be written equivalently as
\be\label{eq:AdS4Metric2}
  ds^2 = - \frac{\Delta_r \Delta_\theta}{B \Xi^2} dt^2 + B\, \textrm{sin}^2 \theta (d\phi + f\, dt)^2 + W \left(\frac{dr^2}{\Delta_r} + \frac{d\theta^2}{\Delta_\theta} \right)\, ,
\ee
with the factors $B$ and $f$ given by
\begin{align}
\begin{split}
  B & \equiv \frac{(a^2 + r_1 r_2)^2 \Delta_\theta - a^2\, \textrm{sin}^2 (\theta)\, \Delta_r}{W \Xi^2}\, ,\\
  f & \equiv \frac{a \Xi \left(\Delta_r - \Delta_\theta (a^2 + r_1 r_2) \right)}{\Delta_\theta (a^2 + r_1 r_2)^2 - a^2 \Delta_r\, \textrm{sin}^2 \theta}\, .
\end{split}
\end{align}
The $\frac{1}{4}$-BPS supersymmetric solutions can be obtained by imposing the condition
\be
  e^{2 \delta_1 + 2 \delta_2} = 1 + \frac{2}{a g}\, ,
\ee
or equivalently,
\be\label{eq:AdS4SUSYCond}
  a = a_0\, ,\quad \textrm{ with}\quad a_0 \equiv \frac{2}{g \left(e^{2 \delta_1 + 2 \delta_2} - 1 \right)}\, .
\ee
The extremal black hole solutions are achieved when the function $\Delta_r (r)$ has a double root, or equivalently when the discriminant of $\Delta_r (r)$ vanishes, which can be viewed as an equation for $m$. We can solve when the discriminant is zero and obtain the extremal value of $m$ as a function of $a$ and $\delta_{1, 2}$, i.e.,
\be\label{eq:AdS4ExtCond}
  m = m_{\text{ext}} (a,\, \delta_1,\, \delta_2)\, .
\ee 
Since this computation is straightforward, we omit the lengthy expression of $m_{\text{ext}} (a,\, \delta_1,\, \delta_2)$. In this case, for $m < m_{\text{ext}}$ the function $\Delta_r (r)$ has two different real roots corresponding to the outer and the inner horizons of a non-extremal black hole. For $m > m_{\text{ext}}$ the function $\Delta_r (r)$ does not have real roots, which implies that the solution has a naked singularity instead of a black hole. We would like to emphasize that the asymptotically AdS$_4$ Kerr-Newman black holes have also been discussed in \cite{Caldarelli:1999xj}. However, the supersymmetric solutions considered \cite{Caldarelli:1999xj} have $\frac{1}{2}$-BPS supersymmetry instead of $\frac{1}{4}$-BPS supersymmetry discussed in \cite{Chong:2004na, Cvetic:2005zi}, which makes some features of the black holes different.

\begin{figure}[!htb]
\begin{center}
  \includegraphics[width=0.6\textwidth]{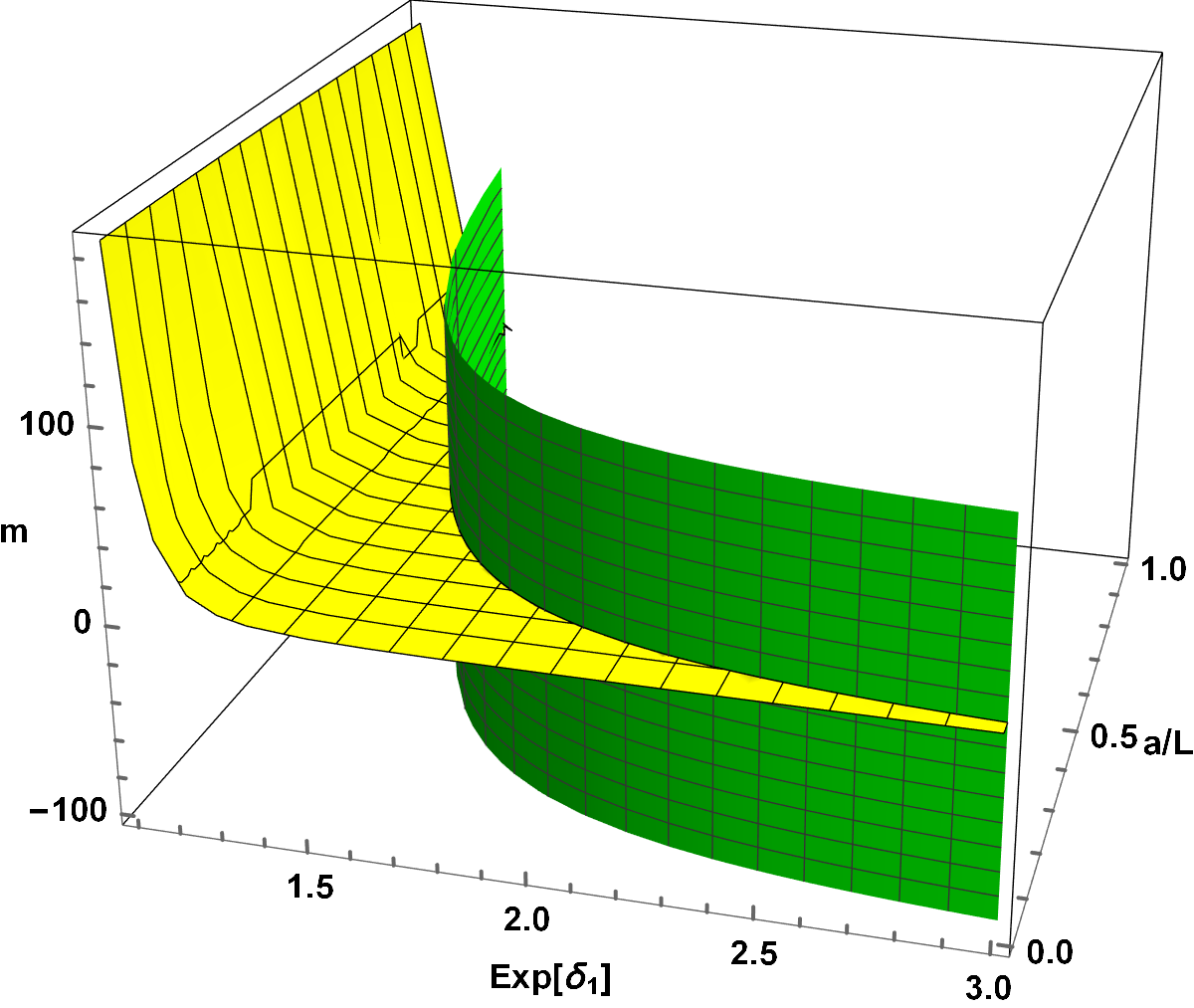}
  \caption{The extremal surface (yellow) and the supersymmetric surface (green)}
   \label{fig:ParameterSurfaces}
\end{center}
\end{figure}

Before moving on, we want to emphasize the parameter space we are exploring. Particularly, we want to focus on the parameters $m,a$ and $\delta_1$ as well as the conditions that characterize supersymmetry and extremality of the black hole. Our goal is to have a black hole solution that is slightly perturbed from these two conditions. Hence, it satisfies both the supersymmetric condition \eqref{eq:AdS4SUSYCond} and the extremal condition \eqref{eq:AdS4ExtCond}. Under the supersymmetric condition \eqref{eq:AdS4SUSYCond}, the extremal condition \eqref{eq:AdS4ExtCond} is equivalent to
\be\label{eq:AdS4RegCond}
  (m g)^2 = \frac{\textrm{cosh}^2 (\delta_1 + \delta_2)}{e^{\delta_1 + \delta_2}\, \textrm{sinh}^3 (\delta_1 + \delta_2)\, \textrm{sinh} (2 \delta_1)\, \textrm{sinh} (2 \delta_2)}\, ,
\ee
which can also be obtained by requiring the black hole solution to have a regular horizon. The two conditions \eqref{eq:AdS4SUSYCond} and \eqref{eq:AdS4RegCond} in \cite{Cvetic:2005zi} contain typos, which have been corrected in \cite{Chow:2013gba, Choi:2018fdc} and also \cite{Cassani:2019mms}. With these two constraints, there are only two independent parameters for asymptotically AdS$_4$ electrically charged rotating BPS black holes. To illustrate the relations of the parameters, we plot in Fig.~\ref{fig:ParameterSurfaces} the codimension-1 supersymmetric surface defined by \eqref{eq:AdS4SUSYCond} together with the codimension-1 extremal surface defined by \eqref{eq:AdS4ExtCond} in the parameter space $(m, a, \delta_1, \delta_2)$, where for simplicity we set $\delta_2 = \delta_1$ and $L = 1$. The intersection of these codimension-1 surfaces is a codimension-2 surface corresponding to the BPS solutions.

We now collect useful properties of the black hole, including the position of the outer horizon in the BPS limit
\be\label{eq:AdS4 r0}
  r_0 = \frac{2 m\, \textrm{sinh} (\delta_1)\, \textrm{sinh} (\delta_2)}{\textrm{cosh} (\delta_1 + \delta_2)}\, ,
\ee
which coincides with the BPS inner horizon. For the thermodynamic quantities of the non-extremal asymptotically AdS$_4$ black holes, the gravitational angular velocity $\Omega$ and the temperature $T_H$ are given by
\be\label{eq:HawkingTemperature}
  \Omega = \frac{a (1 + g^2 r_1 r_2)}{r_1 r_2 + a^2}\, ,\quad T_H = \frac{\Delta'_r}{4 \pi (r_1 r_2 + a^2)}\, ,
\ee
which are evaluated at the outer horizon $r_+$, and the prime ($'$) denotes the derivative with respect to $r$. The other thermodynamic quantities are \cite{Cvetic:2005zi}
\begin{align}
\begin{split}\label{eq:AdS4thermo}
  S & = \frac{\pi (r_1 r_2 + a^2)}{\Xi}\, ,\\
  J & = \frac{m a}{2 \Xi^2} \left(\textrm{cosh} (2 \delta_1) + \textrm{cosh} (2 \delta_2) \right)\, ,\\
  Q_1 = Q_3 & = \frac{m}{4 \Xi}\, \textrm{sinh} (2 \delta_1)\, ,\\
  Q_2 = Q_4 & = \frac{m}{4 \Xi}\, \textrm{sinh} (2 \delta_2)\, .
\end{split}
\end{align}


\subsection{Near-Extremal AdS$_4$ Black Hole Entropy from Gravity Solution}\label{sec:SBH from Gravity}

The asymptotically AdS$_4$ black hole solutions discussed in the previous subsection are in general non-extremal. Since our focus is on near-extremality, we perturb the BPS black hole solution. More precisely, we expand the non-extremal AdS$_4$ black hole solutions around the BPS solution by turning on a small temperature.

We shall do this by studying the parameter space. Before imposing the constraints \eqref{eq:AdS4SUSYCond} and \eqref{eq:AdS4RegCond}, there are 4 parameters that characterize the black hole solution, and we interchange one of these parameters, $a$, with the outer horizon $r_{+}$, where $r_+$ is the biggest root of the equation $\Delta_r (r_+) = 0$, i.e.,
\be \label{rplusequation}
  r_+^2 + a^2 - 2 m r_+ + g^2 \prod_{i=1}^2 \left(r_+ + 2 m\, \textrm{sinh}^2 (\delta_i) \right)\, \Big[\prod_{i=1}^2 \left(r_+ + 2 m\, \textrm{sinh}^2 (\delta_i) \right) + a^2 \Big] = 0\, .
\ee
Note that we only do this replacement of $a$ with $r_{+}$ when we compute the entropy in this subsection. There are two reasons why we make this change. The first is pragmatic: this simplifies the algebra significantly. The second is that the outer horizon $r_+$ plays a clear role in the nAttractor mechanism \cite{Larsen:2018iou}, which will also be relevant for the discussions later in this subsection. Now, we use \eqref{rplusequation} to solve for the parameter $a$ in terms of $r_+$, and the 4 independent parameters for the non-extremal AdS$_4$ black hole solutions are $(r_+, m, \delta_1, \delta_2)$. Correspondingly, there are 4 independent physical quantities $(T_H, J, Q_1, Q_2)$, where we have set $Q_1 = Q_3$ and $Q_2 = Q_4$ as in \eqref{eq:AdS4thermo}. Without loss of generality, we further set $\delta_2 = \delta_1$, and therefore $Q_1=Q_2$, to simplify the discussion.

The black hole entropy in \eqref{eq:AdS4thermo} is valid for any temperature, including small temperature. This is achieved by expanding around the BPS value of the entropy, leading to the expression
\be\label{eq:SBHfromGravity1}
  S = S_* + \left(\frac{C}{T_H} \right)_*\, T_H + \mathcal{O} (T_H^2)\, ,
\ee
where $S_*$ denotes the AdS$_4$ black hole entropy \eqref{eq:AdS4thermo} in the BPS limit
\be\label{eq:SBHext}
  S_* = \frac{2 \pi}{g^2 \left(e^{4\, \delta_1} - 3 \right)}\, ,
\ee
while $C$ is the heat capacity which is linear in $T_H$, and $\left(\frac{C}{T_H} \right)_*$ is evaluated in the BPS limit. Computing $\left(\frac{C}{T_H} \right)_*$ is straightforward
\be
  \left(\frac{C}{T_H} \right)_* = \left(\frac{dS}{dT_H}\right)_* = \left(\frac{\partial S}{\partial r_+}\right)_* \left(\frac{\partial r_+}{\partial T_H}\right)_* + \left(\frac{\partial S}{\partial m}\right)_* \left(\frac{\partial m}{\partial T_H}\right)_* + \left(\frac{\partial S}{\partial \delta_1}\right)_* \left(\frac{\partial \delta_1}{\partial T_H}\right)_*\, ,
\ee
where $\frac{\partial r_+}{\partial T_H}$, $\frac{\partial m}{\partial T_H}$ and $\frac{\partial \delta_1}{\partial T_H}$ can be obtained by inverting the matrix
\be
  \frac{\partial (T_H, J, Q_1)}{\partial (r_+, m, \delta_1)}\, .
\ee
Once the dust settles, the result is
\be\label{eq:SBHfromGravity2}
  \left(\frac{C}{T_H} \right)_* = \frac{8 \sqrt{2} \pi^2 \left(e^{4\, \delta_1} - 1 \right)^{\frac{3}{2}}}{g^3 \left(e^{4\, \delta_1} - 3 \right) \left(e^{8\, \delta_1} + 10\, e^{4\, \delta_1} - 7 \right)}\, .
\ee
We comment that this result can also be obtained by only varying $S$ with respect to $r_+$, i.e.,
\be
  \left(\frac{C}{T_H} \right)_* = \left(\frac{\partial S}{\partial r_+}\right)_* \left(\frac{\partial r_+}{\partial T_H}\right)_*\, .
\ee
This is similar to the AdS$_5$ case discussed in \cite{Larsen:2019oll}, which is related to the nAttractor mechanism \cite{Larsen:2018iou}. This hints that the nAttractor mechanism extends to other dimensions.


\subsection{AdS$_4$ Black Hole Solution in the Near-Horizon Limit}

In this subsection, we consider near-extremal asymptotically AdS$_4$ black holes close to the $\frac{1}{4}$-BPS solutions by introducing a small positive temperature $T$, and discuss the corresponding metrics.

It was discussed in \cite{Caldarelli:1999xj} that for asymptotically AdS$_4$ black holes from the extremal case to non-extremal configurations corresponds to perturbing the parameter $m$ from its extremal value $m_{\text{ext}}$. As we can see from Fig.~\ref{fig:ParameterSurfaces}, when perturbing around the AdS$_4$ $\frac{1}{4}$-BPS black holes, we can deviate from the extremal surface but still stay in the supersymmetric surface by imposing the supersymmetric condition \eqref{eq:AdS4SUSYCond}. Meanwhile, we expand the parameter $m$ around its BPS value given by \eqref{eq:AdS4RegCond} with a small dimensionless parameter $\lambda$ \cite{Matsuo:2010ut} corresponding to near-extremal AdS$_4$ black hole solutions, i.e.,
\be\label{eq:AdS4 NearBPScond}
  m = m_0 (1 + \lambda^2 \widetilde{m})\, ,
\ee
where
\be
  m_0 \equiv \frac{\textrm{cosh} (\delta_1 + \delta_2)}{g\, e^{(\delta_1 + \delta_2)/2}\, \textrm{sinh}^{3/2} (\delta_1 + \delta_2)\, \sqrt{\textrm{sinh} (2 \delta_1)\, \textrm{sinh} (2 \delta_2)}}\, ,
\ee
which coincides with $m_{\textrm{ext}}$ under the supersymmetric condition \eqref{eq:AdS4SUSYCond}. In general $m_{\textrm{ext}} \neq m_0$. A similar limit was also used in \cite{Silva:2006xv} to study the near-BPS black holes and compared with other limits in \cite{Cabo-Bizet:2018ehj}. To summarize, near-extremal AdS$_4$ black holes can be achieved by perturbing the parameter $m$ around $m_{\text{ext}}$ while keeping the other parameters fixed. This is made explicit in this paper by imposing the near-extremal condition \eqref{eq:AdS4 NearBPScond} with the parameter $a$ fixed by the supersymmetric condition \eqref{eq:AdS4SUSYCond}.

Moreover, we perform a near-horizon scaling to the asymptotically AdS$_4$ black hole metric \eqref{eq:AdS4Metric2}, which was first introduced by Bardeen and Horowitz in \cite{Bardeen:1999px} and extensively studied \cite{David:2020ems} for the BPS AdS$_4$ black holes
\be\label{eq:AdS4Scaling}
  r \to r_0 + \lambda\, \widetilde{r}\, ,\quad t \to \frac{\widetilde{t}}{\lambda}\, ,\quad \phi \to \widetilde{\phi} - g \big[\textrm{coth} (2 \delta_1) - 2 \big] \frac{\widetilde{t}}{\lambda}\, .
\ee
In principle, for near-extremal black holes we should consider the near-horizon scaling $r \to r_+ + \lambda\, \widetilde{r}$. However, the near-extremal condition \eqref{eq:AdS4 NearBPScond} implies that $r_+$ and $r_0$ only differ by a constant of order $\lambda$. Hence, we can absorb that constant into $\widetilde{r}$ and still take $r \to r_0 + \lambda\, \widetilde{r}$ in the near-horizon scaling. This kind of near-horizon scaling for near-extremal black holes has been used in \cite{Chen:2010bh}. To summarize, we impose the near-horizon scaling \eqref{eq:AdS4Scaling} together with the condition conditions \eqref{eq:AdS4 NearBPScond} and \eqref{eq:AdS4SUSYCond}.

Taking the limit $\lambda \to 0$, the metric \eqref{eq:AdS4Metric2} becomes
\begin{align}
  ds^2 & = - \frac{\left(e^{8\, \delta_1} + 10\, e^{4\, \delta_1} - 7 \right)\, \left(e^{4\, \delta_1} + \textrm{cos} (2 \theta) \right)}{2\, (e^{4\, \delta_1} + 1)^2}\, g^2\, \widetilde{r}^2\,d\widetilde{t}^2 + \frac{2\, \left(e^{4\, \delta_1} + \textrm{cos} (2 \theta) \right)}{g^2\, \left(e^{8\, \delta_1} + 10\, e^{4\, \delta_1} - 7 \right)}\, \frac{d\widetilde{r}^2}{\widetilde{r}^2} \nonumber\\
  {} & + \frac{2\, \left(e^{4\, \delta_1} + \textrm{cos} (2 \theta) \right)}{g^2\, \left(e^{8\, \delta_1} - 2\, e^{4\, \delta_1} - 1 - 2\, \textrm{cos} (2\, \theta) \right)}\, d\theta^2 \nonumber\\
  {} & \quad + \Lambda_{\textrm{AdS}_4} (\theta) \left[d\widetilde{\phi} + \frac{g^2\, e^{\delta_1}\, (e^{4\, \delta_1} - 3) \sqrt{\textrm{csch} (2\, \delta_1)}\, \textrm{sech} (2\, \delta_1)}{1 + \textrm{coth} (2\, \delta_1)}\, \widetilde{r}\, d\widetilde{t} \right]^2\, ,\label{eq:AdS4 NH metric}
\end{align}
where
\begin{align}
\begin{split}\label{eq:DefAdS4Factors}
  \Lambda_{\textrm{AdS}_4} (\theta) & \equiv \frac{2\, \left(e^{8\, \delta_1} - 2\, e^{4\, \delta_1} - 1 - 2\, \textrm{cos} (2\, \theta) \right)\, \textrm{sin}^2 (\theta)}{g^2\, (e^{4\, \delta_1} - 3)^2\, \left(e^{4\, \delta_1} + \textrm{cos} (2 \theta) \right)}\, .
\end{split}
\end{align}
From the near-horizon, we are now in a position to extract the necessary details to compute the entropy via the Kerr/CFT correspondence. As we shall see, there are several methods to compute the central charges, which requires a rewriting of the near-horizon geometry in different coordinate systems. To make things clearer, we summarize each of these different expressions of the near-horizon metric. The change of coordinates
\be
  \tau \equiv \frac{g^2\, \left(e^{8\, \delta_1} + 10\, e^{4\, \delta_1} - 7 \right)}{2 (e^{4 \delta_1} + 1)}\, \widetilde{t}\, ,\quad \rho \equiv \widetilde{r}\, ,
\ee
allows us to write \eqref{eq:AdS4 NH metric} in Poincar\'e coodinates
\begin{align}
  ds^2 & = \frac{2\, \left(e^{4\, \delta_1} + \textrm{cos} (2 \theta) \right)}{g^2\, \left(e^{8\, \delta_1} + 10\, e^{4\, \delta_1} - 7 \right)}\, \left(- \rho^2\, d\tau^2 + \frac{d\rho^2}{\rho^2} \right) + \frac{2\, \left(e^{4\, \delta_1} + \textrm{cos} (2 \theta) \right)}{g^2\, \left(e^{8\, \delta_1} - 2\, e^{4\, \delta_1} - 1 - 2\, \textrm{cos} (2\, \theta) \right)}\, d\theta^2 \nonumber\\
  {} & \quad + \Lambda_{\textrm{AdS}_4} (\theta) \left[d\widetilde{\phi} + \frac{2 \left(e^{8\, \delta_1} - 4\, e^{4\, \delta_1} + 3 \right) \sqrt{\textrm{csch} (2\, \delta_1)}}{e^{\delta_1} \left(e^{8\, \delta_1} + 10\, e^{4\, \delta_1} - 7 \right)}\, \rho\, d\tau \right]^2\, .\label{eq:AdS4 NH metric Poincare}
\end{align}
Therefore, it is clear that the near-horizon scaling we applied to the metric leaves us with a circle fibered over AdS$_2$, yielding a warped AdS$_3$ geometry. We now see that the near-horizon metric in Poincar\'e coodinates \eqref{eq:AdS4 NH metric Poincare} is in the standard form
\be\label{eq:AdS4 NHmetric standard form}
  ds^2 = f_0 (\theta) \left(- \rho^2\, d\tau^2 + \frac{d\rho^2}{\rho^2} \right) + f_\theta (\theta)\, d\theta^2 + \gamma_{ij} (\theta)\, \left(dx^i + k^i \rho\, d\tau \right)\, \left(dx^j + k^j \rho\, d\tau \right)\, ,
\ee
with $x^i \in \{\widetilde{\phi}\}$ for the AdS$_4$ case, and the coefficients $f_0 (\theta)$, $f_\theta (\theta)$, $k^i$ and $\gamma_{ij} (\theta)$ are functions of $\theta$ in general.

Now, we transform the Poincar\'e coordinates $(\tau, \rho, \theta, \widetilde{\phi})$ in the metric \eqref{eq:AdS4 NH metric Poincare} to the global coordinates $(\hat{t}, \hat{r}, \theta, \hat{\phi})$ using the following relations
\be\label{eq:PoincareTo Global 1}
  g\, \rho = \hat{r} + \sqrt{1 + \hat{r}^2}\, \textrm{cos} (\hat{t})\, ,\quad g^{-1}\, \tau = \frac{\sqrt{1 + \hat{r}^2}\, \textrm{sin} (\hat{t})}{\hat{r} + \sqrt{1 + \hat{r}^2}\, \textrm{cos} (\hat{t})}\, ,
\ee
which leads to
\begin{align}
\begin{split}\label{eq:PoincareTo Global 2}
  - \rho^2\, d\tau^2 + \frac{d\rho^2}{\rho^2} & = - (1 + \hat{r}^2)\, d\hat{t}^2 + \frac{d\hat{r}^2}{1 + \hat{r}^2}\, ,\\
  \rho\, d\tau & = \hat{r}\, d\hat{t} + d\kappa\, ,
\end{split}
\end{align}
where
\be\label{eq:PoincareTo Global 3}
  \kappa \equiv \textrm{log} \left(\frac{1 + \sqrt{1 + \hat{r}^2}\, \textrm{sin} (\hat{t})}{\textrm{cos} (\hat{t}) + \hat{r}\, \textrm{sin} (\hat{t})} \right)\, .
\ee
Consequently, the metric \eqref{eq:AdS4 NH metric Poincare} can be rewritten as
\begin{align}
  ds^2 & = \frac{2\, \left(e^{4\, \delta_1} + \textrm{cos} (2 \theta) \right)}{g^2\, \left(e^{8\, \delta_1} + 10\, e^{4\, \delta_1} - 7 \right)}\, \left[- (1 + \hat{r}^2)\, d\hat{t}^2 + \frac{d\hat{r}^2}{1 + \hat{r}^2} \right] + \frac{2\, \left(e^{4\, \delta_1} + \textrm{cos} (2 \theta) \right)}{g^2\, \left(e^{8\, \delta_1} - 2\, e^{4\, \delta_1} - 1 - 2\, \textrm{cos} (2\, \theta) \right)}\, d\theta^2 \nonumber\\
  {} & \quad + \Lambda_{\textrm{AdS}_4} (\theta) \left[d\hat{\phi} + \frac{2 \left(e^{8\, \delta_1} - 4\, e^{4\, \delta_1} + 3 \right) \sqrt{\textrm{csch} (2\, \delta_1)}}{e^{\delta_1} \left(e^{8\, \delta_1} + 10\, e^{4\, \delta_1} - 7 \right)}\, \hat{r}\, d\hat{t} \right]^2\, ,\label{eq:AdS4 NH metric Global}
\end{align}
where
\be
  \hat{\phi} \equiv \widetilde{\phi} + \frac{2 \left(e^{8\, \delta_1} - 4\, e^{4\, \delta_1} + 3 \right) \sqrt{\textrm{csch} (2\, \delta_1)}}{e^{\delta_1} \left(e^{8\, \delta_1} + 10\, e^{4\, \delta_1} - 7 \right)}\, \kappa\, .
\ee

Besides the near-horizon scaling \eqref{eq:AdS4Scaling}, we can also apply a light-cone scaling in the near-horizon region \cite{Matsuo:2010ut}
\be
  x^+ \equiv \epsilon \left(\phi + \frac{e^{4\, \delta_1} - 3}{e^{4\, \delta_1} - 1}\, g t\right)\, ,\quad x^- \equiv \phi - \frac{e^{4\, \delta_1} - 3}{e^{4\, \delta_1} - 1}\, g t\, ,
\ee
and then consider the following near-horizon scaling in the light-cone coordinates
\be\label{eq:AdS4 LC scaling}
  r \to r_0 + \epsilon\, \widetilde{r}\, ,\quad t \to \frac{e^{4\, \delta_1} - 1}{e^{4\, \delta_1} - 3}\, \frac{x^+ - \epsilon x^-}{2 g \epsilon}\, ,\quad \phi \to \frac{x^+ + \epsilon x^-}{2 \epsilon}\, .
\ee
Together with the condition \eqref{eq:AdS4 NearBPScond} and taking the limit $\epsilon \to 0$, we obtain the near-horizon metric for the AdS$_4$ near-extremal black holes in the coordinates $(x^+,\, \widetilde{r},\, \theta,\, x^-)$
\begin{align}
  ds^2 & = - \frac{\left(e^{4\, \delta_1} - 1 \right)^2 \left(e^{8\, \delta_1} + 10\, e^{4\, \delta_1} - 7 \right) \left(e^{4\, \delta_1} + \textrm{cos} (2 \theta) \right)}{8\, \left(e^{8\, \delta_1} - 2\, e^{4\, \delta_1} - 3 \right)^2}\, \widetilde{r}^2\, dx^{+ 2} + \frac{2\, \left(e^{4\, \delta_1} + \textrm{cos} (2 \theta) \right)}{g^2 \left(e^{8\, \delta_1} + 10\, e^{4\, \delta_1} - 7 \right)}\, \frac{d\widetilde{r}^2}{\widetilde{r}^2} \nonumber\\
  {} & \quad + \frac{2\, \left(e^{4\, \delta_1} + \textrm{cos} (2 \theta) \right)}{g^2\, \left(e^{8\, \delta_1} - 2\, e^{4\, \delta_1} - 1 - 2\, \textrm{cos} (2\, \theta) \right)}\, d\theta^2 + \Lambda_{\textrm{AdS}_4} (\theta) \left[dx^- + \frac{e^{\delta_1}\, \textrm{sech} (2\, \delta_1)}{\textrm{csch} (2\, \delta_1)^{3/2}}\, g\, \widetilde{r}\, dx^+ \right]^2\, ,\label{eq:AdS4 LightCone Metric}
\end{align}
where $\Lambda_{\textrm{AdS$_4$}} (\theta)$ is the same as \eqref{eq:DefAdS4Factors}. Introducing some new coordinates
\be
  \hat{x}^+ \equiv \frac{g \left(e^{4\, \delta_1} - 1 \right) \left(e^{8\, \delta_1} + 10\, e^{4\, \delta_1} - 7 \right)}{4\, \left(e^{8\, \delta_1} - 2\, e^{4\, \delta_1} - 3 \right)}\, x^+\, ,\quad \hat{\rho} \equiv \widetilde{r}\, ,\quad \hat{x}^- \equiv x^-\, ,
\ee
we can rewrite the near-horizon metric in the light-cone coordinates \eqref{eq:AdS4 LightCone Metric} as
\begin{align}
  ds^2 & = \frac{2\, \left(e^{4\, \delta_1} + \textrm{cos} (2 \theta) \right)}{g^2 \left(e^{8\, \delta_1} + 10\, e^{4\, \delta_1} - 7 \right)} \left[- \hat{\rho}^2\, d\hat{x}^{+2} + \frac{d\hat{\rho}^2}{\hat{\rho}^2} \right] + \frac{2\, \left(e^{4\, \delta_1} + \textrm{cos} (2 \theta) \right)}{g^2\, \left(e^{8\, \delta_1} - 2\, e^{4\, \delta_1} - 1 - 2\, \textrm{cos} (2\, \theta) \right)}\, d\theta^2 \nonumber\\
  {} & \quad + \Lambda_{\textrm{AdS}_4} (\theta) \left[d\hat{x}^- + \frac{2 \left(e^{8\, \delta_1} - 4\, e^{4\, \delta_1} + 3 \right) \sqrt{\textrm{csch} (2\, \delta_1)}}{e^{\delta_1} \left(e^{8\, \delta_1} + 10\, e^{4\, \delta_1} - 7 \right)}\, \hat{\rho}\, d\hat{x}^+ \right]^2\, .\label{eq:AdS4 LightCone Metric Poincare}
\end{align}
We see that the metric \eqref{eq:AdS4 LightCone Metric Poincare} is in the standard form
\be\label{eq:AdS4 NHmetric LightCone standard form}
  ds^2 = f_0 (\theta) \left(- \hat{\rho}^2\, d\hat{x}^{+2} + \frac{d\hat{\rho}^2}{\hat{\rho}^2} \right) + f_\theta (\theta)\, d\theta^2 + \gamma_{ij} (\theta)\, \left(dx^i + k^i \hat{\rho}\, d\hat{x}^+ \right)\, \left(dx^j + k^j \hat{\rho}\, d\hat{x}^+ \right)\, ,
\ee
with $x^i \in \{\hat{x}^- \}$ for the AdS$_4$ case, and $k^i$, $f_0 (\theta)$, $f_\theta (\theta)$ and $\gamma_{ij} (\theta)$ remain the same as \eqref{eq:AdS4 NHmetric standard form}. To summarize, we now have several different expressions for the near-horizon metric in Poincar\'e and global coordinates. This is useful when we utilize the near-extremal Kerr/CFT correspondence.


\subsection{Near-Extremal AdS$_4$ Black Hole Entropy from Cardy Formula}\label{sec:CardyFormula}

After obtaining the various expressions of the near-horizon metric of the asymptotically AdS$_4$ black holes, we are now ready to compute the central charges and the Frolov-Thorne temperatures using the near-extremal Kerr/CFT correspondence as well as hidden conformal symmetry of the near-horizon geometry to find the AdS$_4$ black hole entropy in the near-extremal limit. For the left central charge $c_L$ and the right central charge $c_R$, there are two different ways for computing each of them, depending on which coordinate system we choose. We summarize each of these diverse approaches as a consistency check on our computation as well as to keep things self-contained.

The Kerr/CFT correspondence was originally posed for asymptotically flat extremal Kerr black holes \cite{Guica:2008mu} and was later shown to also be valid for asymptotically AdS black holes \cite{Lu:2008jk, Chow:2008dp}. For the near-extremal case, \cite{Chen:2010bh, Matsuo:2010ut} initiated some progress and we extend those results here by computing the entropies of near-extremal AdS$_4$ black holes via the Cardy formula. Before further exploring the near-extremal case, let us take a step back and recall how the Kerr/CFT correspondence works. The basic idea is the following. Taking the Bardeen-Horowitz near-horizon scaling \cite{Bardeen:1999px}, the near-horizon geometry of an asymptotically flat or asymptotically AdS extremal black hole contains $U(1)$ cycles fibered on AdS$_2$. The near-horizon asymptotic symmetries are characterized by diffeomorphims generated by the vectors
\be
  \zeta_\epsilon = \epsilon (\phi) \frac{\partial}{\partial \phi} - r\, \epsilon' (\phi) \frac{\partial}{\partial r}\, .
\ee
The mode expansion of a diffeomorphism generating vector $\zeta$ is
\be
  \zeta_{(n)} = - e^{- i n \widetilde{\phi}} \frac{\partial}{\partial \widetilde{\phi}} - i n r e^{- i n \widetilde{\phi}} \frac{\partial}{\partial \hat{r}}\, .
\ee
We can define a 2-form $k_\zeta$ for a general perturbation $h_{\mu\nu}$ around the background metric $g_{\mu\nu}$ as
\begin{align}
  k_\zeta [h,\, g] & \equiv - \frac{1}{4} \epsilon_{\alpha\beta\mu\nu} \Big[\zeta^\nu D^\mu h - \zeta^\nu D_\sigma h^{\mu\sigma} + \zeta_\sigma D^\nu h^{\mu\sigma} + \frac{1}{2} h D^\nu \zeta^\mu - h^{\nu\sigma} D_\sigma \zeta^\mu \nonumber\\
  {} & \qquad\qquad\quad + \frac{1}{2} h^{\sigma\nu} (D^\mu \zeta_\sigma + D_\sigma \zeta^\mu)\Big] dx^\alpha \wedge dx^\beta\, .
\end{align}
We also define the Lie derivative with respect to $\zeta$, denoted by $\mathcal{L}_\zeta$, as
\be
  \mathcal{L}_\zeta g_{\mu\nu} \equiv \zeta^\rho \partial_\rho g_{\mu\nu} + g_{\rho\nu} \partial_\mu \zeta^\rho + g_{\mu\rho} \partial_\nu \zeta^\rho\, .
\ee
The left central charge $c_L$ of the near-horizon Virasoro algebra can be computed using the Kerr/CFT correspondence in two slightly different ways. For the first method, the central charge can be computed using the following integral \cite{Guica:2008mu, Lu:2008jk, Chow:2008dp}
\be\label{eq:cL First Approach}
  \frac{1}{8 \pi G} \int_{\partial \Sigma} k_{\zeta_{(m)}} [\mathcal{L}_{\zeta_{(n)}} g,\, g] = - \frac{i}{12} c_L\, (m^3 + \alpha m)\, \delta_{m+n,\, 0}\, ,
\ee
where $g$ denotes the near-horizon metric of the near-extremal AdS$_4$ black hole in global coordinates \eqref{eq:AdS4 NH metric Global}. An explicit evaluation of \eqref{eq:cL First Approach} shows that
\be\label{eq:AdS4 VirasoroIntegral cL 1}
  c_L = \frac{24 \sqrt{2 \left(e^{4\, \delta_1} - 1 \right)}}{g^2 \left(e^{8 \delta_1} + 10\, e^{4\, \delta_1} - 7 \right)}\, .
\ee
The other way of computing $c_L$ is to evaluate the following integral \cite{Matsuo:2010ut}
\be
  \frac{1}{8 \pi G_N} \int_{\partial \Sigma} k_{\xi_n} \left[\mathcal{L}_{\xi_m} \bar{g},\, \bar{g} \right] = \delta_{n+m,\, 0}\, n^3 \frac{c_L}{12}\, ,
\ee
where $\bar{g}$ denotes the standard form \eqref{eq:AdS4 NHmetric standard form} of the near-horizon metric of the near-extremal AdS$_4$ black hole in Poincar\'e coordinates \eqref{eq:AdS4 NH metric Poincare}. More precisely, we obtain with the unit $G_N = 1$
\be\label{eq:AdS4 VirasoroIntegral cL 2}
  c_L = \frac{3 k_{\widetilde{\phi}}}{G_N} \int_0^{\pi} d\theta\, \sqrt{\textrm{Det} \left(\gamma_{ij} (\theta) \right)\, f_\theta (\theta)} = \frac{24 \sqrt{2 \left(e^{4\, \delta_1} - 1 \right)}}{g^2 \left(e^{8 \delta_1} + 10\, e^{4\, \delta_1} - 7 \right)}\, ,
\ee
which matches exactly the result of $c_L$ \eqref{eq:AdS4 VirasoroIntegral cL 1} from the first approach.

The right central charge $c_R$ can also be obtained in two different ways. Although we describe the two methods, we prefer one method over the other because of its robustness. The first approach is to compute the quasi-local charge \cite{Brown:1992br, Balasubramanian:1999re, Matsuo:2010ut} using the standard form of the near-horizon metric of the near-extremal AdS$_4$ black hole in Poincar\'e coordinates \eqref{eq:AdS4 NHmetric standard form}, which is given by the integral
\be\label{eq:cR First Approach}
  \frac{c_R}{12} = \frac{1}{8 \pi G_N} \int dx^i\, d\theta\, \frac{k_i k_j \gamma_{ij} (\theta)\, \sqrt{\textrm{Det} \left(\gamma_{ij} (\theta) \right)\, f_\theta (\theta)}}{2 \Lambda_0 f_0 (\theta)}\, ,
\ee
where $f_0 (\theta)$, $f_\theta (\theta)$, $\gamma_{ij} (\theta)$ and $k_i$ are defined in \eqref{eq:AdS4 NHmetric standard form}, and the parameter $\Lambda_0$ denotes a UV cutoff in $r$. This approach has been used to compute  the right central charge $c_R$ for near-extremal AdS$_5$ black holes \cite{Nian:2020qsk}. For the four-dimensional case, the integral \eqref{eq:cR First Approach} can be applied to the near-horizon metric \eqref{eq:AdS4 NH metric Poincare} in Poincar\'e coordinates to compute $c_R$. However, the result is not very illuminating due to the unfixed cutoff $\Lambda_0$.

To compute $c_R$, we choose a more concrete approach using light-cone coordinates as introduced in \cite{Matsuo:2010ut}. More precisely, a scale-covariant right central charge $c_R^{(cov)}$ can be computed from the near-horizon metric \eqref{eq:AdS4 LightCone Metric Poincare} by using
\be\label{eq:AdS4 VirasoroIntegral cR 2}
  c_R^{(cov)} = 3 k_-\, \epsilon \int_0^\pi d\theta\, \sqrt{\textrm{Det} \left(\gamma_{ij} (\theta) \right)\, f_\theta (\theta)} = \frac{24\, \epsilon \sqrt{2 \left(e^{4\, \delta_1} - 1 \right)}}{g^2 \left(e^{8 \delta_1} + 10\, e^{4\, \delta_1} - 7 \right)}\, ,
\ee
where the factors $\gamma_{ij} (\theta)$, $f_\theta (\theta)$ and $k_-$ are defined in \eqref{eq:AdS4 NHmetric LightCone standard form}. Like in \cite{Matsuo:2010ut}, we can define a scale-invariant right central charge $c_R \equiv c_R^{(cov)} / \epsilon$, which in this case is
\be
  c_R = \frac{24 \sqrt{2 \left(e^{4\, \delta_1} - 1 \right)}}{g^2 \left(e^{8 \delta_1} + 10\, e^{4\, \delta_1} - 7 \right)}\, .
\ee
We see that the result is exactly the same as the left central charge computed in \eqref{eq:AdS4 VirasoroIntegral cL 1} and \eqref{eq:AdS4 VirasoroIntegral cL 2}. To summarize, the explicit expression for the integral changes slightly depending on the coordinate system, and we have shown that all the results do indeed lead to the same central charge.

Now that we have taken care of the central charges, and have consistently gotten that $c_{L}=c_{R}$, the final ingredient is the Frolov-Thorne temperatures $T_L$ and $T_R$. We have seen in \cite{David:2020ems} that for the BPS case $T_R = 0$. For the near-extremal case, $T_L$ can still be computed in the same way discussed in \cite{David:2020ems}, and its value remains the same as the BPS case, as it is unaffected by whether we impose the condition \eqref{eq:AdS4 NearBPScond}. Therefore, we find
\be \label{TL}
  T_L = \frac{e^{\delta_1} \left(e^{8\, \delta_1} + 10\, e^{4\, \delta_1} - 7 \right) \sqrt{\textrm{sinh} (2\, \delta_1)}}{4 \pi \left(e^{8\, \delta_1} - 4\, e^{4\, \delta_1} + 3 \right)}\, .
\ee
On the other hand, $T_R$ is proportional to the physical Hawking temperature $T_H$. To find the exact expression of $T_R$, we apply the technique of hidden conformal symmetry. This method was first introduced in \cite{Castro:2010fd}, and later generalized to many different cases. The basic idea is to define a set of near-horizon conformal coordinates and corresponding locally-defined vector fields with $SU(2, \mathbb{R})$ Lie algebra, such that the wave equation of an uncharged massless scalar field becomes the quadratic Casimir of the $SU(2, \mathbb{R})$ Lie algebra. In this way, we can fix the Frolov-Thorne temperatures $T_{L, R}$ and the mode numbers $N_{L, R}$ for non-extremal black holes.

In particular, \cite{Chen:2010bh} has considered the hidden conformal symmetry of an AdS$_4$ black hole close to the solutions discussed in this paper. We can apply the same technique by first expanding $\Delta_r$ defined in \eqref{eq:AdS4MetricFactors}
\be\label{eq:LeadingDelta}
  \Delta_r = k (r - r_+) (r - r_s) + \mathcal{O} \left( (r - r_+)^3 \right)\, ,
\ee
where $k$ and $r_s$ can be read off from the Taylor expansion to quadratic order in $(r - r_+)$. Based on hidden conformal symmetry \cite{Chen:2010bh}, the right temperature is 
\be\label{eq:TRfromHiddenConformal}
  T_R = \frac{k (r_+ - r_s)}{4 \pi a \Xi}\, .
\ee
Applying the definition the Hawking temperature $T_H$ \eqref{eq:HawkingTemperature} to the leading order expression of $\Delta_r$ \eqref{eq:LeadingDelta}, we find for $b=a$ that
\be\label{eq:LeadingTH}
  T_H = \frac{k (r_+ - r_s)}{4 \pi (r_1^2 + a^2)}\, .
\ee
Combining \eqref{eq:TRfromHiddenConformal} with \eqref{eq:LeadingTH}, we obtain
\be
  T_R = \frac{r_1^2 + a^2}{a \Xi}\, T_H\, .
\ee
We also find that the expression obtained using hidden conformal symmetry for $T_{L}$ is \eqref{TL} as expected. Using the Cardy formula, we obtain the near-extremal AdS$_4$ black hole entropy
\begin{align}
  S & = \frac{\pi^2}{3} c_L T_L + \frac{\pi^2}{3} c_R T_R \nonumber\\
  {} & = S_* + \left(\frac{C}{T_H} \right)_* T_H\, ,\nonumber\\
  {} & = \frac{2 \pi}{g^2 \left(e^{4\, \delta_1} - 3 \right)} +  \frac{8 \sqrt{2} \pi^2 \left(e^{4\, \delta_1} - 1 \right)^{\frac{3}{2}}}{g^3 \left(e^{4\, \delta_1} - 3 \right) \left(e^{8\, \delta_1} + 10\, e^{4\, \delta_1} - 7 \right)}\, T_H\, ,\label{eq:SBH from Kerr CFT}
\end{align}
where the BPS entropy $S_*$ is
\be\label{eq:SBH from Kerr CFT 1}
  S_* = \frac{\pi^2}{3} c_L T_L\, ,
\ee
while the near-extremal correction to the black hole entropy is
\be\label{eq:SBH from Kerr CFT 2}
  \delta S = \frac{\pi^2}{3} c_R T_R \equiv \left(\frac{C}{T_H} \right)_* T_H\, .
\ee
We see that this result from the near-horizon CFT$_2$ and the Cardy formula is exactly the same as the results from the gravity side (\eqref{eq:SBHfromGravity1}, \eqref{eq:SBHext} and \eqref{eq:SBHfromGravity2}).


\subsection{Near-Extremal AdS$_4$ Black Hole Entropy from Boundary CFT}

What remains is the computation of the near-extremal entropy from the boundary CFT. In the BPS limit, the AdS$_4$ black hole entropy can be obtained by extremizing an entropy function, which was derived by the superconformal index or supersymmetric localization of the 3d ABJM theory on the boundary of electrically charged rotating AdS$_4$ BPS black holes \cite{Choi:2019zpz, Nian:2019pxj}. More precisely, the BPS entropy function is
\be\label{eq:EntropyFct-2}
  S (\widetilde{\Delta}_I,\, \widetilde{\omega}) = - \frac{4 \sqrt{2}\, i\, k^{\frac{1}{2}} N^{\frac{3}{2}}}{3} \frac{\sqrt{\widetilde{\Delta}_1 \widetilde{\Delta}_2 \widetilde{\Delta}_3 \widetilde{\Delta}_4}}{\widetilde{\omega}} + \widetilde{\omega} J + \sum_I \widetilde{\Delta}_I Q_I + \Lambda \left(\sum_I \widetilde{\Delta}_I - \widetilde{\omega} - 2 \pi i \right)\, ,
\ee
where $\widetilde{\Delta}_I$ are chemical potentials corresponding to the electric charges $Q_I$, and $\widetilde{\omega}$ is the angular velocity. To extremize the entropy function \eqref{eq:EntropyFct-2}, we solve the equations
\be
  \frac{\partial S}{\partial \widetilde{\Delta}_I} = 0\, ,\quad \frac{\partial S}{\partial \widetilde{\omega}} = 0\, ,
\ee
which can be expressed explicitly as
\begin{align}
  Q_I + \Lambda & = \frac{4 \sqrt{2}\, i\, k^{\frac{1}{2}} N^{\frac{3}{2}}}{3} \frac{\sqrt{\widetilde{\Delta}_1 \widetilde{\Delta}_2 \widetilde{\Delta}_3 \widetilde{\Delta}_4}}{2 \widetilde{\Delta}_I \widetilde{\omega}}\, ,\label{eq:ExtEq1}\\
  J - \Lambda & = - \frac{4 \sqrt{2}\, i\, k^{\frac{1}{2}} N^{\frac{3}{2}}}{3} \frac{\sqrt{\widetilde{\Delta}_1 \widetilde{\Delta}_2 \widetilde{\Delta}_3 \widetilde{\Delta}_4}}{\tilde{\omega}^2}\, .\label{eq:ExtEq2}
\end{align}
Substituting these equations back into the entropy function \eqref{eq:EntropyFct-2}, we obtain
\be\label{eq:S_Interm}
  S = - 2 \pi i \Lambda\, .
\ee
Moreover, the equations \eqref{eq:ExtEq1} and \eqref{eq:ExtEq2} can be combined into one equation:
\begin{align}\label{eq:ExtEq3}
  {} & \, Q_1 Q_2 Q_3 Q_4 + \Lambda \left(\sum_{I < J < K} Q_I Q_J Q_K \right) + \Lambda^2 \left(\sum_{I < J} Q_I Q_J \right) + \Lambda^3 \left(\sum_I Q_I \right) + \Lambda^4 \nonumber\\
  = & \, - \frac{2}{9} k N^3 (\Lambda^2 - 2 \Lambda J + J^2)\, ,
\end{align}
which can be written more compactly as
\be\label{eq:ExtEq4}
  \Lambda^4 + A\, \Lambda^3 + B\, \Lambda^2 + C\, \Lambda + D = 0\, ,
\ee
with the real-valued coefficients
\begin{align}
\begin{split}
  A & = \sum_{I=1}^4 Q_I\, ,\\
  B & = \sum_{I < J} Q_I Q_J + \frac{2}{9} k N^3\, ,\\
  C & = \sum_{I < J < K} Q_I Q_J Q_K - \frac{4}{9} k N^3 J\, ,\\
  D & = Q_1 Q_2 Q_3 Q_4 + \frac{2}{9} k N^3 J^2\, .
\end{split}
\end{align}
In order to obtain a real-valued black hole entropy, the expression \eqref{eq:S_Interm} implies that $\Lambda$ should have a purely imaginary root. Since \eqref{eq:ExtEq4} is a quartic equation of $\Lambda$ with real coefficients, the imaginary roots should come in pairs. Consequently, \eqref{eq:ExtEq4} can be factorized as
\be\label{eq:ExtEq5}
  (\Lambda^2 + \alpha) (\Lambda^2 + \beta\, \Lambda + \mu) = \Lambda^4 + \beta\, \Lambda^3 + (\alpha + \mu)\, \Lambda^2 + \alpha \beta\, \Lambda + \alpha \mu\, .
\ee
Comparing \eqref{eq:ExtEq5} with \eqref{eq:ExtEq4}, we find
\be\label{eq:Coeff1}
  A = \beta\, ,\quad B = \alpha + \mu\, ,\quad C = \alpha \beta\, ,\quad D = \alpha \mu\, ,
\ee
or equivalently,
\be\label{eq:Coeff2}
  \alpha = \frac{C}{A}\, ,\quad \beta = A\, ,\quad \mu = B - \frac{C}{A} = \frac{A D}{C}\, .
\ee
According to \eqref{eq:S_Interm}, the imaginary root $\Lambda = i \sqrt{\alpha} = i \sqrt{\frac{C}{A}}$ leads to the real-valued AdS$_4$ BPS black hole entropy
\be
  S_{BH}^* = 2 \pi \sqrt{\frac{Q_1 Q_2 Q_3 + Q_1 Q_2 Q_4 + Q_1 Q_3 Q_4 + Q_2 Q_3 Q_4 - \frac{4}{9} k N^3 J}{Q_1 + Q_2 + Q_3 + Q_4}}\, .
\ee
For the special case $Q_1 = Q_3$, $Q_2 = Q_4$, the expression above becomes
\be\label{eq:BPS SBH from bdy CFT 1}
  S_{BH}^* = \frac{2 \pi}{3} \sqrt{\frac{9 Q_1 Q_2 (Q_1 + Q_2) - 2 k J N^3}{Q_1 + Q_2}}\, .
\ee
After imposing the identifications of parameters introduced in \cite{Choi:2018fdc, Choi:2019zpz, Nian:2019pxj}
\be\label{eq:IdentifyParameters}
  Q_{BH,I} = \frac{g}{2} Q_I\, ,\quad J_{BH} = J\, , \qquad I \in \{1,\cdots ,4\}
\ee
and using an entry from the AdS/CFT dictionary
\be\label{eq:AdSCFTdict}
  \frac{1}{G_N} = \frac{2 \sqrt{2}}{3} g^2 k^{\frac{1}{2}} N^{\frac{3}{2}}\, ,
\ee
we can rewrite the BPS black hole entropy \eqref{eq:BPS SBH from bdy CFT 1} as
\be
  S_{BH}^* = \frac{\pi}{g^2 G} \frac{J_{BH}}{ \left( \frac{2}{g} Q_{BH, 1} + \frac{2}{g} Q_{BH, 2} \right)}\, ,\label{eq:BPS SBH from bdy CFT 2}
\ee
which can be subsequently written in terms of the free parameters $(\delta_1, \delta_2)$ on the gravity side in the BPS limit. For the special case $\delta_1 = \delta_2$, the BPS black hole entropy obtained from the boundary CFT is
\be\label{eq:BPS SBH from bdy CFT 3}
  S_{BH}^* = \frac{2 \pi}{g^2 \left(e^{4\, \delta_1} - 3 \right)}\, ,
\ee
which is exactly the same as the BPS result from the gravity side \eqref{eq:SBHext} and the one from the near-horizon Kerr/CFT correspondence \eqref{eq:SBH from Kerr CFT 1}.

In addition to the black hole entropy, the electric charges $Q_I$'s and the angular momentum $J$ should also satisfy a constraint, which originates from the consistency of two expressions of $\mu$ in \eqref{eq:Coeff2}, i.e.,
\be
  B - \frac{C}{A} - \frac{A D}{C} = 0\, .
\ee
More explicitly, for the special case $Q_1 = Q_3$, $Q_2 = Q_4$ the constraint is
\be
  \frac{2}{9} k N^3 + (Q_1 + Q_2)^2 + \frac{2 k J N^3}{9 (Q_1 + Q_2)} + \frac{2 k J N^3 \left[Q_1 Q_2 + J (Q_1 + Q_2) \right]}{2 k J N^3 - 9 Q_1 Q_2 (Q_1 + Q_2)} = 0\, .
\ee
We emphasize that the constraint is not unique. A constraint multiplied by a constant or some regular function of $Q_{I}$ and $J$ can produce new constraints. For later convenience, we define
\be\label{eq:Choose h}
  h \equiv \frac{J^2}{4 g^5 (Q_1 + Q_2)^2} \left[\frac{2}{9} k N^3 + (Q_1 + Q_2)^2 + \frac{2 k J N^3}{9 (Q_1 + Q_2)} + \frac{2 k J N^3 \left[Q_1 Q_2 + J (Q_1 + Q_2) \right]}{2 k J N^3 - 9 Q_1 Q_2 (Q_1 + Q_2)} \right]\, ,
\ee
whose BPS value will be called $h_*$, and
\be
  h_* = 0
\ee
is one of the BPS constraints. So far we have only considered the BPS black holes from the boundary CFT in this subsection. To extend the BPS results to the near-extremal case, similar to the AdS$_5$ case discussed in \cite{Nian:2020qsk}, we generalize the quartic equation \eqref{eq:ExtEq4} from the BPS limit to the near-extremal case by perturbing $\Lambda$ and $h$ as
\be
  (\Lambda + \delta \Lambda)^4 + A\, (\Lambda + \delta \Lambda)^3 + B\, (\Lambda + \delta \Lambda)^2 + C\, (\Lambda + \delta \Lambda) + D + (h_* + \delta h) = 0\, ,
\ee
which at the order $\mathcal{O} (\delta \Lambda)$ is
\be\label{eq:PerturbQuarticEq}
  (4 \Lambda^3 + 3 A\, \Lambda^2 + 2 B\, \Lambda + C)\, \delta \Lambda + \delta h = 0\, .
\ee
For the root $\Lambda = i \sqrt{\frac{C}{A}}$, which has led to the BPS black hole entropy, we can solve \eqref{eq:PerturbQuarticEq} and obtain
\be
  \delta \Lambda = \frac{\delta h}{2 C - 2 i \sqrt{\frac{C}{A}} \left(B - 2 \frac{C}{A} \right) }\, .
\ee
Based on \eqref{eq:S_Interm}, the correction to the BPS black hole entropy is
\be\label{eq:dS_Interm}
  \delta S = - 2 \pi i\, \delta \Lambda\, .
\ee
Hence, only the imaginary part of $\delta \Lambda$ will contribute to the real part of $\delta S$. If we assume that $\delta h$ is purely imaginary, then
\begin{align}
  \textrm{Im} (\delta \Lambda) & = \delta h\, \textrm{Re} \left[\frac{1}{2 C - 2 i \sqrt{\frac{C}{A}} \left(B - 2 \frac{C}{A} \right) } \right] \nonumber\\
  {} & = \frac{2 C\, \delta h}{\left[2 C - 2 i \sqrt{\frac{C}{A}} \left(B - 2 \frac{C}{A} \right)\right]\, \left[2 C + 2 i \sqrt{\frac{C}{A}} \left(B - 2 \frac{C}{A} \right)\right] } \nonumber\\
  {} & = \frac{\delta h}{2 C + \frac{2}{A} \left(B - 2 \frac{C}{A} \right)^2}\, .
\end{align}
Therefore, for real-valued $\delta S$ we have
\be
  \delta S = - 2 \pi i\, \textrm{Im} (\delta \Lambda) = \frac{- \pi i\, \delta h}{C + \frac{1}{A} \left(B - 2 \frac{C}{A} \right)^2}\, .
\ee
We view $\delta h$ as a small change of $h$ from its BPS value, i.e.,
\be
  \delta h = h - h_* = h\, .
\ee
We can compute $\delta h$ by
\be\label{eq:dh}
  \delta h = \frac{\partial h}{\partial Q_I}\, \delta Q_I + \frac{\partial h}{\partial J}\, \delta J\, ,
\ee
with the transformations similar to the AdS$_5$ case \cite{Nian:2020qsk}
\be
  \delta Q_I = \eta Q_I\, ,\quad \delta J_i = \eta J_i\, .
\ee
For the near-extremal case, we relate the transformation parameter $\eta$ with the temperature change
\be
  2 \pi i\, \delta T_H = 2 \eta\, ,\label{eq:dT and lambda}
\ee
where $\delta T_H = T_H - T_H^* = T_H$. Now, we apply \eqref{eq:dh} to the explicit choice of $h$ given by \eqref{eq:Choose h}. In the unit $G_N = 1$, the near-extremal correction to the BPS entropy for the special case $\delta_1 = \delta_2$ becomes
\be\label{eq:dS from bdy CFT}
  \delta S = \frac{8 \sqrt{2} \pi^2 \left(e^{4\, \delta_1} - 1 \right)^{\frac{3}{2}}}{g^3 \left(e^{4\, \delta_1} - 3 \right)\, \left(e^{8\, \delta_1} + 10\, e^{4\, \delta_1} - 7 \right)}\, T_H \equiv \left(\frac{C}{T_H} \right)_*\, T_H\, .
\ee
Combining the BPS black hole entropy from the boundary CFT \eqref{eq:BPS SBH from bdy CFT 3} and the near-extremal correction \eqref{eq:dS from bdy CFT}, we obtain the near-extremal AdS$_4$ black hole entropy from the boundary CFT
\begin{align}
  S_{BH} & = S_{BH}^* + \delta S \nonumber\\
  {} & = \frac{2 \pi}{g^2 \left(e^{4\, \delta_1} - 3 \right)} + \frac{8 \sqrt{2} \pi^2 \left(e^{4\, \delta_1} - 1 \right)^{\frac{3}{2}}}{g^3 \left(e^{4\, \delta_1} - 3 \right)\, \left(e^{8\, \delta_1} + 10\, e^{4\, \delta_1} - 7 \right)}\, T_H \nonumber\\
  {} & \equiv S_* + \left(\frac{C}{T_H} \right)_*\, T_H\, ,
\end{align}
which matches perfectly with the results from gravity solution (\eqref{eq:SBHfromGravity1}, \eqref{eq:SBHext} and \eqref{eq:SBHfromGravity2}) and from the near-horizon Kerr/CFT correspondence (\eqref{eq:SBH from Kerr CFT}, \eqref{eq:SBH from Kerr CFT 1} and \eqref{eq:SBH from Kerr CFT 2}).


\section{Hawking Radiation and Near-Extremal AdS$_4$ Black Hole}\label{sec:HawkingRadiation}

In Sec.~\ref{sec:AdS4}, we have derived the near-extremal AdS$_4$ black hole entropy using three different approaches and obtained one universal result. In particular, the approach of the near-horizon Kerr/CFT correspondence shows that there exists a near-horizon CFT$_2$, which accounts for the low-energy spectrum of the black hole microstates.

As we have seen in Subsection~\ref{sec:CardyFormula}, the near-extremal black hole entropy can be decomposed into the contributions from the left and the right sectors of the near-horizon CFT$_2$. The expression from the canonical ensemble is
\be\label{eq:SBH canonical}
  S_{BH} = \frac{\pi^2}{3} c_L T_L + \frac{\pi^2}{3} c_R T_R \, ,
\ee
which has been discussed extensively for the asymptotically flat black holes \cite{Cvetic:1996kv, Larsen:1997ge, Cvetic:1997uw, Cvetic:1997xv, Cvetic:1997vp, Larsen:1999pp, Cvetic:2010mn, Castro:2012av}, while the expression from the microcanonical ensemble is \cite{Cardy:1986ie}
\be\label{eq:SBH microcanonical}
  S_{BH} = 2 \pi \sqrt{\frac{c_L N_L}{6}} + 2 \pi \sqrt{\frac{c_R N_R}{6}}\, ,
\ee
where $N_L$ and $N_R$ are the left and the right mode numbers, respectively. The relation \eqref{eq:SBH microcanonical} can be applied to asymptotically AdS black holes as well (see e.g. \cite{Jing:2002aq}). Comparing the expressions \eqref{eq:SBH canonical} and \eqref{eq:SBH microcanonical}, we find that the temperatures $T_{L, R}$ can be related to the mode numbers $N_{L, R}$
\be\label{eq:TL TR NL NR}
  T_L = \frac{1}{\pi} \sqrt{\frac{6 N_L}{c_L}}\, ,\quad T_R = \frac{1}{\pi} \sqrt{\frac{6 N_R}{c_R}}\, .
\ee
The explicit expressions of $N_L$ and $N_R$ for near-extremal AdS$_4$ black holes considered in this paper with $\delta_1 = \delta_2$ are
\begin{align}
\begin{split}\label{eq:NLNR}
  N_L & = \frac{e^{8\, \delta_1} + 10\, e^{4\, \delta_1} - 7}{4 \sqrt{2} g^2 \left(e^{4\, \delta_1} - 3 \right)^2 \sqrt{e^{4\, \delta_1} - 1}}\, ,\\
  N_R & = \frac{4 \sqrt{2} \pi^2 \left(e^{4\, \delta_1} - 1 \right)^{\frac{5}{2}} T_H^2}{g^4 \left(e^{4\, \delta_1} - 3 \right)^2\, \left(e^{8\, \delta_1} + 10\, e^{4\, \delta_1} - 7 \right)}\, .
\end{split}
\end{align}
Suppose that the left and the right mode numbers in the BPS limit are $N_L^*$ and $N_R^*$ respectively, where
\be
  N_R^* = 0\, .
\ee
As discussed in \cite{Callan:1996dv, Nian:2020qsk}, for the near-extremal case the left-moving and the right-moving modes become
\begin{align}
\begin{split}
  N_L & = N_L^* + \delta N_L \approx N_L^*\, ,\\
  N_R & = N_R^* + \delta N_R = \delta N_R\, ,
\end{split}
\end{align}
with $\delta N_L = \delta N_R \ll N_L^*$, which can be seen from \eqref{eq:NLNR}. Because $T_H / g \ll 1$ for near-extremal AdS$_4$ black holes, the right mode number $N_R \sim (T_H / g)^2$ is much smaller than the left mode number $N_L \sim (T_H / g)^0$. If we assume that the right modes obey a microcanonical ensemble, then the partition function of the right sector can be written as
\be
  Z_R = \sum_{N_R} q^{N_R} d(N_R) = \sum_{N_R} q^{N_R} e^{S_R} = \sum_{N_R} q^{N_R} e^{2 \pi \sqrt{c_R N_R / 6}}\, ,
\ee
where $q \equiv e^{- 1 / T_R}$. We evaluate this partition function using a saddle-point approximation with respect to $N_R$, and the result is
\be\label{eq:log(q)}
  \delta N_R = N_R = q \frac{\partial}{\partial q} \textrm{log} Z_R \approx \frac{c_R \pi^2}{6\, \left(\textrm{log} (q) \right)^2}\, ,\quad \textrm{with}\quad \textrm{log} (q) < 0\, .
\ee
The inequality $\textrm{log} (q) < 0$ can be justified by the definition $q \equiv e^{- 1 / T_R}$. For near-extremal AdS$_4$ black holes $0 < T_R \ll 1$, which implies that $0 < q \ll 1$ and consequently $\textrm{log} (q) < 0$. The occupation number in the right sector is given by Bose-Einstein statistics
\be\label{eq:Rel q and T}
  \rho_R (k_0) = \frac{q^n}{1 - q^n} = \frac{e^{- \frac{k_0}{T_R}}}{1- e^{- \frac{k_0}{T_R}}}\, ,
\ee
where $n$ is the momentum quantum number of the mode moving in the time circle for the near-horizon region of AdS$_4$ black holes, and $k_0$ is the typical energy of the right-moving modes. From \eqref{eq:log(q)} we can solve for $q$ in terms of $\delta N_R = N_R$, and then combining it with \eqref{eq:Rel q and T} we obtain
\be\label{eq:TR New}
  T_R = \frac{k_0}{\pi n} \sqrt{6\, \frac{\delta N_R}{c_R}} = \frac{1}{\pi} \sqrt{\frac{6\, N_R}{c_R}}\, ,
\ee
where we used $k_0 = n$. A similar expression holds for $T_L$, i.e.,
\be\label{eq:TL New}
  T_L = \frac{1}{\pi} \sqrt{\frac{6\, N_L}{c_L}}\, .
\ee
We see that \eqref{eq:TR New} and \eqref{eq:TL New} are completely consistent with \eqref{eq:TL TR NL NR}. Since the right sector obeys the Bose-Einstein statistics, the typical energy $k_0$ of the right modes can be characterized by the right temperature $T_R$. In the limit $k_0 \sim T_R \ll T_L$, the occupation number in the left sector can be approximated as
\be\label{eq: rhoL approx}
  \rho_L (k_0) = \frac{e^{- \frac{k_0}{T_L}}}{1- e^{- \frac{k_0}{T_L}}} \approx \frac{T_L}{k_0} = \frac{1}{\pi k_0} \sqrt{\frac{6\, N_L}{c_L}}\, .
\ee

According to \cite{Callan:1996dv, Nian:2020qsk}, Hawking radiation can be formulated as a scattering process of left and right modes in the near-horizon CFT$_2$. Therefore, we can evaluate the Hawking radiation rate for near-extremal AdS$_4$ black holes based on the analyses above
\be
  d\Gamma \sim \frac{d^4 k}{k_0} \frac{1}{p_0^L\, p_0^R} |\mathcal{A}|^2\, c_L\, \rho_L (k_0)\, \rho_R (k_0)\, ,
\ee
where the central charge $c_L$ provides the degrees of freedom for a given momentum quantum number $n$, and $\mathcal{A}$ is the disc amplitude of strings depending on details of the near-horizon CFT$_2$. From \eqref{eq: rhoL approx} we see that
\be
  c_L\, \rho_L (k_0) \sim S_L \,\propto\, (\textrm{horizon area})\, .
\ee
Consequently, the Hawking radiation rate becomes
\be\label{eq:HawkingRadRate}
  d\Gamma \sim (\textrm{horizon area})\cdot \frac{e^{- \frac{k_0}{T_R}}}{1- e^{- \frac{k_0}{T_R}}} d^4 k\, ,
\ee
which is a consequence of Bose-Einstein statistics, and implies that the radiation spectrum is thermal and governed by a temperature $T_R$ proportional to the Hawking temperature $T_H$. Therefore, we have found a microscopic formalism of Hawking radiation in the near-horizon CFT$_2$. According to this picture, the scattering of modes is unitary;  hence there is no information loss during the Hawking radiation process.

Since the boundary CFT can exactly reproduce the near-extremal black hole entropy \eqref{eq:SBH canonical}, this microscopic formalism of Hawking radiation can in principle be embedded in higher-dimensional boundary CFT, which is the 3d superconformal ABJM theory for AdS$_4$ black holes.

Like the AdS$_5$ case discussed in \cite{Nian:2020qsk}, we have not taken into account the global structure of AdS space. Particularly, due to the conformal boundary of AdS space, once the radiation reaches the boundary, it will bounce back and head towards the black hole. Therefore, our current model provides a microscopic description for the Hawking radiation immediately after creation. We leave the full evolution of Hawking radiation for future work.

\section{Discussion}\label{sec:Discussion}

We have studied the electrically charged rotating AdS$_4$ black holes in the near-extremal limit. Moreover, by studying the parameter space we have successfully defined a way to approach near-extremal supersymmetric black holes. We have then computed the entropy using three different approaches: (i) from the gravity solution, (ii) from the near-horizon CFT$_2$ via the Kerr/CFT correspondence and (iii) from the boundary CFT via the AdS/CFT correspondence. Remarkably, these three results match precisely, giving us a universal and unique expression for the entropy in the near-extremal limit. This supports the near-extremal microstate counting in the boundary CFT and in the near-horizon CFT$_2$. We also have shown that the extension of the Kerr/CFT correspondence, originally posed for extremality, to near-extremal black holes is valid. Using the results of near-extremal black hole entropy, we provide a microscopic description of Hawking radiation, and qualitatively show that unitarity and information are preserved during the Hawking radiation process.

The success of this work provides motivation to further study near-extremality in other dimensions and indeed show that the three diverse entropy computations lead to one universal entropy. Besides the near-extremal AdS$_5$ black holes discussed in \cite{Nian:2020qsk} and the AdS$_4$ case discussed in this paper, we can also consider the known AdS$_6$ and AdS$_7$ \cite{Larsen:2020lhg} black hole solutions. Similar results from different approaches listed in Fig.~\ref{fig:M1} are expected. Moreover, the unifying picture Fig.~\ref{fig:M1} can potentially be valid beyond the Bekenstein-Hawking entropy. Hence, it would be interesting to study the subleading corrections to the Bekenstein-Hawking entropy and see if the different approaches still provide a unique expression for the entropy, in the same spirit of \cite{Nian:2017hac, Liu:2017vll, Jeon:2017aif, Liu:2017vbl, Hristov:2018lod, Liu:2018bac, Gang:2019uay, PandoZayas:2019hdb, Hristov:2019xku, GonzalezLezcano:2020yeb, PandoZayas:2020iqr}. A recent work \cite{Ghosh:2020rwf} shows that Sen's classical entropy function formalism \cite{Sen:2005wa} can be applied to asymptotically AdS$_4$ black holes to capture higher derivative corrections to the Bekenstein-Hawking entropy, which complements the methods in Fig.~\ref{fig:M1}.

Besides the microstate counting of black holes, a more interesting question is how to use field theory techniques to study 
dynamical process in black hole physics. For instance, Hawking radiation on asymptotically AdS black holes has been studied within the framework of AdS/CFT correspondence previously in \cite{Hubeny:2009ru, Hubeny:2009rc}. Some recent progress has been made for microscopic description of Hawking-Page transition \cite{Copetti:2020dil}. Another related problem is to reproduce the Page curve in the black hole evaporation process \cite{Page:1993wv, Page:2013dx, Nian:2019buz}, which has been studied in the framework of 2d JT gravity coupled to a 2d bath CFT \cite{Penington:2019npb, Almheiri:2019hni, Almheiri:2019psf, Almheiri:2019qdq}. Our approach in \cite{Nian:2020qsk} and in this paper provides another powerful framework of studying these problems. In order to do that, however, we have to first carefully study the Hawking radiation at a later time in the dual boundary field theory and in the near-horizon CFT$_2$ to resolve the issues from the global property of AdS space. We hope to refine our microscopic models and study these more physical problems in the near future.

\section*{Acknowledgements}

We would like to thank Leo A. Pando Zayas for collaboration in the early stages of this project and for many enlightening discussions. We also would like to thank Alfredo Gonz\'alez Lezcano for many helpful discussions. This work was supported in part by the U.S. Department of Energy under grant DE-SC0007859. M.D. is supported by the NSF Graduate Research Fellowship Program under NSF Grant Number: DGE 1256260. J.N. would like to thank Korea Institute for Advanced Study (KIAS), Kavli Institute for the Physics and Mathematics of the Universe (Kavli IPMU) and Simons Center for Geometry and Physics (SCGP) for warm hospitality during various stages of this project.

\bibliography{NearExtKerrCFT}

\providecommand{\href}[2]{#2}\begingroup\raggedright\begin{thebibliography}{10}

\bibitem{Strominger:1996sh}
A.~Strominger and C.~Vafa, ``{Microscopic origin of the Bekenstein-Hawking
  entropy},'' \href{http://dx.doi.org/10.1016/0370-2693(96)00345-0}{{\em Phys.
  Lett.} {\bfseries B379} (1996) 99--104},
\href{http://arxiv.org/abs/hep-th/9601029}{{\ttfamily arXiv:hep-th/9601029
  [hep-th]}}.

\bibitem{Maldacena:1997re}
J.~M. Maldacena, ``{The Large N limit of superconformal field theories and
  supergravity},'' \href{http://dx.doi.org/10.1023/A:1026654312961,
  10.4310/ATMP.1998.v2.n2.a1}{{\em Int. J. Theor. Phys.} {\bfseries 38} (1999)
  1113--1133}, \href{http://arxiv.org/abs/hep-th/9711200}{{\ttfamily
  arXiv:hep-th/9711200 [hep-th]}}.
[Adv. Theor. Math. Phys.2,231(1998)].

\bibitem{Witten:1998qj}
E.~Witten, ``{Anti-de Sitter space and holography},''
  \href{http://dx.doi.org/10.4310/ATMP.1998.v2.n2.a2}{{\em Adv. Theor. Math.
  Phys.} {\bfseries 2} (1998) 253--291},
\href{http://arxiv.org/abs/hep-th/9802150}{{\ttfamily arXiv:hep-th/9802150
  [hep-th]}}.

\bibitem{Benini:2015eyy}
F.~Benini, K.~Hristov, and A.~Zaffaroni, ``{Black hole microstates in AdS$_{4}$
  from supersymmetric localization},''
  \href{http://dx.doi.org/10.1007/JHEP05(2016)054}{{\em JHEP} {\bfseries 05}
  (2016) 054},
\href{http://arxiv.org/abs/1511.04085}{{\ttfamily arXiv:1511.04085 [hep-th]}}.

\bibitem{Cabo-Bizet:2018ehj}
A.~Cabo-Bizet, D.~Cassani, D.~Martelli, and S.~Murthy, ``{Microscopic origin of
  the Bekenstein-Hawking entropy of supersymmetric AdS$_{5}$ black holes},''
  \href{http://dx.doi.org/10.1007/JHEP10(2019)062}{{\em JHEP} {\bfseries 10}
  (2019) 062}, \href{http://arxiv.org/abs/1810.11442}{{\ttfamily
  arXiv:1810.11442 [hep-th]}}.

\bibitem{Choi:2018hmj}
S.~Choi, J.~Kim, S.~Kim, and J.~Nahmgoong, ``{Large AdS black holes from
  QFT},''
\href{http://arxiv.org/abs/1810.12067}{{\ttfamily arXiv:1810.12067 [hep-th]}}.

\bibitem{Benini:2018ywd}
F.~Benini and P.~Milan, ``{Black Holes in 4D $\mathcal{N}$=4 Super-Yang-Mills
  Field Theory},'' \href{http://dx.doi.org/10.1103/PhysRevX.10.021037}{{\em
  Phys. Rev. X} {\bfseries 10} no.~2, (2020) 021037},
  \href{http://arxiv.org/abs/1812.09613}{{\ttfamily arXiv:1812.09613
  [hep-th]}}.

\bibitem{Choi:2019miv}
S.~Choi and S.~Kim, ``{Large AdS$_6$ black holes from CFT$_5$},''
\href{http://arxiv.org/abs/1904.01164}{{\ttfamily arXiv:1904.01164 [hep-th]}}.

\bibitem{Kantor:2019lfo}
G.~K\'antor, C.~Papageorgakis, and P.~Richmond, ``{AdS$_{7}$ black-hole entropy
  and 5D $ \mathcal{N} $ = 2 Yang-Mills},''
  \href{http://dx.doi.org/10.1007/JHEP01(2020)017}{{\em JHEP} {\bfseries 01}
  (2020) 017}, \href{http://arxiv.org/abs/1907.02923}{{\ttfamily
  arXiv:1907.02923 [hep-th]}}.

\bibitem{Nahmgoong:2019hko}
J.~Nahmgoong, ``{6d superconformal Cardy formulas},''
\href{http://arxiv.org/abs/1907.12582}{{\ttfamily arXiv:1907.12582 [hep-th]}}.

\bibitem{Choi:2019zpz}
S.~Choi, C.~Hwang, and S.~Kim, ``{Quantum vortices, M2-branes and black
  holes},''
\href{http://arxiv.org/abs/1908.02470}{{\ttfamily arXiv:1908.02470 [hep-th]}}.

\bibitem{Nian:2019pxj}
J.~Nian and L.~A. Pando~Zayas, ``{Microscopic entropy of rotating electrically
  charged AdS$_{4}$ black holes from field theory localization},''
  \href{http://dx.doi.org/10.1007/JHEP03(2020)081}{{\em JHEP} {\bfseries 03}
  (2020) 081}, \href{http://arxiv.org/abs/1909.07943}{{\ttfamily
  arXiv:1909.07943 [hep-th]}}.

\bibitem{Cabo-Bizet:2019eaf}
A.~Cabo-Bizet and S.~Murthy, ``{Supersymmetric phases of 4d $N=4$ SYM at large
  $N$},'' \href{http://arxiv.org/abs/1909.09597}{{\ttfamily arXiv:1909.09597
  [hep-th]}}.

\bibitem{Bobev:2019zmz}
N.~Bobev and P.~M. Crichigno, ``{Universal spinning black holes and theories of
  class $ \mathcal{R} $},''
  \href{http://dx.doi.org/10.1007/JHEP12(2019)054}{{\em JHEP} {\bfseries 12}
  (2019) 054}, \href{http://arxiv.org/abs/1909.05873}{{\ttfamily
  arXiv:1909.05873 [hep-th]}}.

\bibitem{Benini:2019dyp}
F.~Benini, D.~Gang, and L.~A. Pando~Zayas, ``{Rotating Black Hole Entropy from
  M5 Branes},'' \href{http://dx.doi.org/10.1007/JHEP03(2020)057}{{\em JHEP}
  {\bfseries 03} (2020) 057}, \href{http://arxiv.org/abs/1909.11612}{{\ttfamily
  arXiv:1909.11612 [hep-th]}}.

\bibitem{Hosseini:2019lkt}
S.~M. Hosseini, K.~Hristov, and A.~Zaffaroni, ``{Microstates of rotating
  AdS$_{5}$ strings},'' \href{http://dx.doi.org/10.1007/JHEP11(2019)090}{{\em
  JHEP} {\bfseries 11} (2019) 090},
  \href{http://arxiv.org/abs/1909.08000}{{\ttfamily arXiv:1909.08000
  [hep-th]}}.

\bibitem{Choi:2019dfu}
S.~Choi and C.~Hwang, ``{Universal 3d Cardy Block and Black Hole Entropy},''
  \href{http://dx.doi.org/10.1007/JHEP03(2020)068}{{\em JHEP} {\bfseries 03}
  (2020) 068}, \href{http://arxiv.org/abs/1911.01448}{{\ttfamily
  arXiv:1911.01448 [hep-th]}}.

\bibitem{Crichigno:2020ouj}
P.~M. Crichigno and D.~Jain, ``{The 5d Superconformal Index at Large $N$ and
  Black Holes},'' \href{http://dx.doi.org/10.1007/JHEP09(2020)124}{{\em JHEP}
  {\bfseries 09} (2020) 124}, \href{http://arxiv.org/abs/2005.00550}{{\ttfamily
  arXiv:2005.00550 [hep-th]}}.

\bibitem{Guica:2008mu}
M.~Guica, T.~Hartman, W.~Song, and A.~Strominger, ``{The Kerr/CFT
  Correspondence},'' \href{http://dx.doi.org/10.1103/PhysRevD.80.124008}{{\em
  Phys. Rev.} {\bfseries D80} (2009) 124008},
\href{http://arxiv.org/abs/0809.4266}{{\ttfamily arXiv:0809.4266 [hep-th]}}.

\bibitem{Lu:2008jk}
H.~Lu, J.~Mei, and C.~N. Pope, ``{Kerr/CFT Correspondence in Diverse
  Dimensions},'' \href{http://dx.doi.org/10.1088/1126-6708/2009/04/054}{{\em
  JHEP} {\bfseries 04} (2009) 054},
\href{http://arxiv.org/abs/0811.2225}{{\ttfamily arXiv:0811.2225 [hep-th]}}.

\bibitem{Chow:2008dp}
D.~D.~K. Chow, M.~Cvetic, H.~Lu, and C.~N. Pope, ``{Extremal Black Hole/CFT
  Correspondence in (Gauged) Supergravities},''
  \href{http://dx.doi.org/10.1103/PhysRevD.79.084018}{{\em Phys. Rev.}
  {\bfseries D79} (2009) 084018},
\href{http://arxiv.org/abs/0812.2918}{{\ttfamily arXiv:0812.2918 [hep-th]}}.

\bibitem{David:2020ems}
M.~David, J.~Nian, and L.~A. Pando~Zayas, ``{Gravitational Cardy Limit and AdS
  Black Hole Entropy},'' \href{http://arxiv.org/abs/2005.10251}{{\ttfamily
  arXiv:2005.10251 [hep-th]}}.

\bibitem{Larsen:2019oll}
F.~Larsen, J.~Nian, and Y.~Zeng, ``{AdS$_{5}$ black hole entropy near the BPS
  limit},'' \href{http://dx.doi.org/10.1007/JHEP06(2020)001}{{\em JHEP}
  {\bfseries 06} (2020) 001}, \href{http://arxiv.org/abs/1907.02505}{{\ttfamily
  arXiv:1907.02505 [hep-th]}}.

\bibitem{Nian:2020qsk}
J.~Nian and L.~A. Pando~Zayas, ``{Toward an effective CFT$_{2}$ from $
  \mathcal{N} $ = 4 super Yang-Mills and aspects of Hawking radiation},''
  \href{http://dx.doi.org/10.1007/JHEP07(2020)120}{{\em JHEP} {\bfseries 07}
  (2020) 120}, \href{http://arxiv.org/abs/2003.02770}{{\ttfamily
  arXiv:2003.02770 [hep-th]}}.

\bibitem{Bardeen:1999px}
J.~M. Bardeen and G.~T. Horowitz, ``{The Extreme Kerr throat geometry: A Vacuum
  analog of AdS(2) x S**2},''
  \href{http://dx.doi.org/10.1103/PhysRevD.60.104030}{{\em Phys. Rev.}
  {\bfseries D60} (1999) 104030},
\href{http://arxiv.org/abs/hep-th/9905099}{{\ttfamily arXiv:hep-th/9905099
  [hep-th]}}.

\bibitem{Callan:1996dv}
C.~G. Callan and J.~M. Maldacena, ``{D-brane approach to black hole quantum
  mechanics},'' \href{http://dx.doi.org/10.1016/0550-3213(96)00225-8}{{\em
  Nucl. Phys.} {\bfseries B472} (1996) 591--610},
\href{http://arxiv.org/abs/hep-th/9602043}{{\ttfamily arXiv:hep-th/9602043
  [hep-th]}}.

\bibitem{Matsuo:2010ut}
Y.~Matsuo and T.~Nishioka, ``{New Near Horizon Limit in Kerr/CFT},''
  \href{http://dx.doi.org/10.1007/JHEP12(2010)073}{{\em JHEP} {\bfseries 12}
  (2010) 073}, \href{http://arxiv.org/abs/1010.4549}{{\ttfamily arXiv:1010.4549
  [hep-th]}}.

\bibitem{Chen:2010bh}
B.~Chen and J.~Long, ``{On Holographic description of the Kerr-Newman-AdS-dS
  black holes},'' \href{http://dx.doi.org/10.1007/JHEP08(2010)065}{{\em JHEP}
  {\bfseries 08} (2010) 065}, \href{http://arxiv.org/abs/1006.0157}{{\ttfamily
  arXiv:1006.0157 [hep-th]}}.

\bibitem{Penington:2019npb}
G.~Penington, ``{Entanglement Wedge Reconstruction and the Information
  Paradox},'' \href{http://dx.doi.org/10.1007/JHEP09(2020)002}{{\em JHEP}
  {\bfseries 09} (2020) 002}, \href{http://arxiv.org/abs/1905.08255}{{\ttfamily
  arXiv:1905.08255 [hep-th]}}.

\bibitem{Almheiri:2019hni}
A.~Almheiri, R.~Mahajan, J.~Maldacena, and Y.~Zhao, ``{The Page curve of
  Hawking radiation from semiclassical geometry},''
  \href{http://dx.doi.org/10.1007/JHEP03(2020)149}{{\em JHEP} {\bfseries 03}
  (2020) 149}, \href{http://arxiv.org/abs/1908.10996}{{\ttfamily
  arXiv:1908.10996 [hep-th]}}.

\bibitem{Almheiri:2019psf}
A.~Almheiri, N.~Engelhardt, D.~Marolf, and H.~Maxfield, ``{The entropy of bulk
  quantum fields and the entanglement wedge of an evaporating black hole},''
  \href{http://dx.doi.org/10.1007/JHEP12(2019)063}{{\em JHEP} {\bfseries 12}
  (2019) 063}, \href{http://arxiv.org/abs/1905.08762}{{\ttfamily
  arXiv:1905.08762 [hep-th]}}.

\bibitem{Almheiri:2019qdq}
A.~Almheiri, T.~Hartman, J.~Maldacena, E.~Shaghoulian, and A.~Tajdini,
  ``{Replica Wormholes and the Entropy of Hawking Radiation},''
  \href{http://dx.doi.org/10.1007/JHEP05(2020)013}{{\em JHEP} {\bfseries 05}
  (2020) 013}, \href{http://arxiv.org/abs/1911.12333}{{\ttfamily
  arXiv:1911.12333 [hep-th]}}.

\bibitem{Larsen:2020lhg}
F.~Larsen and S.~Paranjape, ``{Thermodynamics of Near BPS Black Holes in
  AdS$_4$ and AdS$_7$},'' \href{http://arxiv.org/abs/2010.04359}{{\ttfamily
  arXiv:2010.04359 [hep-th]}}.

\bibitem{Chong:2004na}
Z.~W. Chong, M.~Cvetic, H.~Lu, and C.~N. Pope, ``{Charged rotating black holes
  in four-dimensional gauged and ungauged supergravities},''
  \href{http://dx.doi.org/10.1016/j.nuclphysb.2005.03.034}{{\em Nucl. Phys.}
  {\bfseries B717} (2005) 246--271},
\href{http://arxiv.org/abs/hep-th/0411045}{{\ttfamily arXiv:hep-th/0411045
  [hep-th]}}.

\bibitem{Cvetic:2005zi}
M.~Cvetic, G.~W. Gibbons, H.~Lu, and C.~N. Pope, ``{Rotating black holes in
  gauged supergravities: Thermodynamics, supersymmetric limits, topological
  solitons and time machines},''
\href{http://arxiv.org/abs/hep-th/0504080}{{\ttfamily arXiv:hep-th/0504080
  [hep-th]}}.

\bibitem{Caldarelli:1999xj}
M.~M. Caldarelli, G.~Cognola, and D.~Klemm, ``{Thermodynamics of
  Kerr-Newman-AdS black holes and conformal field theories},''
  \href{http://dx.doi.org/10.1088/0264-9381/17/2/310}{{\em Class. Quant. Grav.}
  {\bfseries 17} (2000) 399--420},
  \href{http://arxiv.org/abs/hep-th/9908022}{{\ttfamily arXiv:hep-th/9908022}}.

\bibitem{Chow:2013gba}
{Chow, David D. K. and Comp\'ere, Geoffrey}, ``{Dyonic AdS black holes in
  maximal gauged supergravity},''
  \href{http://dx.doi.org/10.1103/PhysRevD.89.065003}{{\em Phys. Rev.}
  {\bfseries D89} no.~6, (2014) 065003},
\href{http://arxiv.org/abs/1311.1204}{{\ttfamily arXiv:1311.1204 [hep-th]}}.

\bibitem{Choi:2018fdc}
S.~Choi, C.~Hwang, S.~Kim, and J.~Nahmgoong, ``{Entropy Functions of BPS Black
  Holes in AdS$_{4}$ and AdS$_{6}$},''
  \href{http://dx.doi.org/10.3938/jkps.76.101}{{\em J.\ Korean Phys.\ Soc.}
  {\bfseries 76} no.~2, (2020) 101--108},
  \href{http://arxiv.org/abs/1811.02158}{{\ttfamily arXiv:1811.02158
  [hep-th]}}.

\bibitem{Cassani:2019mms}
D.~Cassani and L.~Papini, ``{The BPS limit of rotating AdS black hole
  thermodynamics},'' \href{http://dx.doi.org/10.1007/JHEP09(2019)079}{{\em
  JHEP} {\bfseries 09} (2019) 079},
  \href{http://arxiv.org/abs/1906.10148}{{\ttfamily arXiv:1906.10148
  [hep-th]}}.

\bibitem{Larsen:2018iou}
F.~Larsen, ``{A nAttractor mechanism for nAdS$_{2}$/nCFT$_{1}$ holography},''
  \href{http://dx.doi.org/10.1007/JHEP04(2019)055}{{\em JHEP} {\bfseries 04}
  (2019) 055}, \href{http://arxiv.org/abs/1806.06330}{{\ttfamily
  arXiv:1806.06330 [hep-th]}}.

\bibitem{Silva:2006xv}
P.~J. Silva, ``{Thermodynamics at the BPS bound for Black Holes in AdS},''
  \href{http://dx.doi.org/10.1088/1126-6708/2006/10/022}{{\em JHEP} {\bfseries
  10} (2006) 022}, \href{http://arxiv.org/abs/hep-th/0607056}{{\ttfamily
  arXiv:hep-th/0607056}}.

\bibitem{Brown:1992br}
J.~D. Brown and J.~W. York, Jr., ``{Quasilocal energy and conserved charges
  derived from the gravitational action},''
  \href{http://dx.doi.org/10.1103/PhysRevD.47.1407}{{\em Phys. Rev.} {\bfseries
  D47} (1993) 1407--1419},
\href{http://arxiv.org/abs/gr-qc/9209012}{{\ttfamily arXiv:gr-qc/9209012
  [gr-qc]}}.

\bibitem{Balasubramanian:1999re}
V.~Balasubramanian and P.~Kraus, ``{A Stress tensor for Anti-de Sitter
  gravity},'' \href{http://dx.doi.org/10.1007/s002200050764}{{\em Commun. Math.
  Phys.} {\bfseries 208} (1999) 413--428},
\href{http://arxiv.org/abs/hep-th/9902121}{{\ttfamily arXiv:hep-th/9902121
  [hep-th]}}.

\bibitem{Castro:2010fd}
A.~Castro, A.~Maloney, and A.~Strominger, ``{Hidden Conformal Symmetry of the
  Kerr Black Hole},'' \href{http://dx.doi.org/10.1103/PhysRevD.82.024008}{{\em
  Phys. Rev. D} {\bfseries 82} (2010) 024008},
  \href{http://arxiv.org/abs/1004.0996}{{\ttfamily arXiv:1004.0996 [hep-th]}}.

\bibitem{Cvetic:1996kv}
M.~Cvetic and D.~Youm, ``{Entropy of nonextreme charged rotating black holes in
  string theory},'' \href{http://dx.doi.org/10.1103/PhysRevD.54.2612}{{\em
  Phys. Rev. D} {\bfseries 54} (1996) 2612--2620},
  \href{http://arxiv.org/abs/hep-th/9603147}{{\ttfamily arXiv:hep-th/9603147}}.

\bibitem{Larsen:1997ge}
F.~Larsen, ``{A String model of black hole microstates},''
  \href{http://dx.doi.org/10.1103/PhysRevD.56.1005}{{\em Phys. Rev. D}
  {\bfseries 56} (1997) 1005--1008},
  \href{http://arxiv.org/abs/hep-th/9702153}{{\ttfamily arXiv:hep-th/9702153}}.

\bibitem{Cvetic:1997uw}
M.~Cvetic and F.~Larsen, ``{General rotating black holes in string theory: Grey
  body factors and event horizons},''
  \href{http://dx.doi.org/10.1103/PhysRevD.56.4994}{{\em Phys. Rev. D}
  {\bfseries 56} (1997) 4994--5007},
  \href{http://arxiv.org/abs/hep-th/9705192}{{\ttfamily arXiv:hep-th/9705192}}.

\bibitem{Cvetic:1997xv}
M.~Cvetic and F.~Larsen, ``{Grey body factors for rotating black holes in
  four-dimensions},''
  \href{http://dx.doi.org/10.1016/S0550-3213(97)00541-5}{{\em Nucl. Phys. B}
  {\bfseries 506} (1997) 107--120},
  \href{http://arxiv.org/abs/hep-th/9706071}{{\ttfamily arXiv:hep-th/9706071}}.

\bibitem{Cvetic:1997vp}
M.~Cvetic and F.~Larsen, ``{Black hole horizons and the thermodynamics of
  strings},'' \href{http://dx.doi.org/10.1016/S0920-5632(97)00685-3}{{\em Nucl.
  Phys. B Proc. Suppl.} {\bfseries 62} (1998) 443--453},
  \href{http://arxiv.org/abs/hep-th/9708090}{{\ttfamily arXiv:hep-th/9708090}}.

\bibitem{Larsen:1999pp}
F.~Larsen, ``{Rotating Kaluza-Klein black holes},''
  \href{http://dx.doi.org/10.1016/S0550-3213(00)00064-X}{{\em Nucl. Phys. B}
  {\bfseries 575} (2000) 211--230},
  \href{http://arxiv.org/abs/hep-th/9909102}{{\ttfamily arXiv:hep-th/9909102}}.

\bibitem{Cvetic:2010mn}
M.~Cvetic, G.~W. Gibbons, and C.~N. Pope, ``{Universal Area Product Formulae
  for Rotating and Charged Black Holes in Four and Higher Dimensions},''
  \href{http://dx.doi.org/10.1103/PhysRevLett.106.121301}{{\em Phys. Rev.
  Lett.} {\bfseries 106} (2011) 121301},
\href{http://arxiv.org/abs/1011.0008}{{\ttfamily arXiv:1011.0008 [hep-th]}}.

\bibitem{Castro:2012av}
A.~Castro and M.~J. Rodriguez, ``{Universal properties and the first law of
  black hole inner mechanics},''
  \href{http://dx.doi.org/10.1103/PhysRevD.86.024008}{{\em Phys. Rev.}
  {\bfseries D86} (2012) 024008},
\href{http://arxiv.org/abs/1204.1284}{{\ttfamily arXiv:1204.1284 [hep-th]}}.

\bibitem{Cardy:1986ie}
J.~L. Cardy, ``{Operator Content of Two-Dimensional Conformally Invariant
  Theories},'' \href{http://dx.doi.org/10.1016/0550-3213(86)90552-3}{{\em Nucl.
  Phys. B} {\bfseries 270} (1986) 186--204}.

\bibitem{Jing:2002aq}
J.-l. Jing, ``{Cardy-Verlinde formula and entropy bounds in Kerr-Newman AdS(4)
  / dS(4) black hole backgrounds},''
  \href{http://dx.doi.org/10.1103/PhysRevD.66.024002}{{\em Phys. Rev. D}
  {\bfseries 66} (2002) 024002},
  \href{http://arxiv.org/abs/hep-th/0201247}{{\ttfamily arXiv:hep-th/0201247}}.

\bibitem{Nian:2017hac}
J.~Nian and X.~Zhang, ``{Entanglement Entropy of ABJM Theory and Entropy of
  Topological Black Hole},''
  \href{http://dx.doi.org/10.1007/JHEP07(2017)096}{{\em JHEP} {\bfseries 07}
  (2017) 096},
\href{http://arxiv.org/abs/1705.01896}{{\ttfamily arXiv:1705.01896 [hep-th]}}.

\bibitem{Liu:2017vll}
J.~T. Liu, L.~A. Pando~Zayas, V.~Rathee, and W.~Zhao, ``{Toward Microstate
  Counting Beyond Large N in Localization and the Dual One-loop Quantum
  Supergravity},'' \href{http://dx.doi.org/10.1007/JHEP01(2018)026}{{\em JHEP}
  {\bfseries 01} (2018) 026},
\href{http://arxiv.org/abs/1707.04197}{{\ttfamily arXiv:1707.04197 [hep-th]}}.

\bibitem{Jeon:2017aif}
I.~Jeon and S.~Lal, ``{Logarithmic Corrections to Entropy of Magnetically
  Charged AdS$_4$ Black Holes},''
  \href{http://dx.doi.org/10.1016/j.physletb.2017.09.026}{{\em Phys. Lett.}
  {\bfseries B774} (2017) 41--45},
\href{http://arxiv.org/abs/1707.04208}{{\ttfamily arXiv:1707.04208 [hep-th]}}.

\bibitem{Liu:2017vbl}
J.~T. Liu, L.~A. Pando~Zayas, V.~Rathee, and W.~Zhao, ``{One-Loop Test of
  Quantum Black Holes in anti-de Sitter Space},''
  \href{http://dx.doi.org/10.1103/PhysRevLett.120.221602}{{\em Phys. Rev.
  Lett.} {\bfseries 120} no.~22, (2018) 221602},
\href{http://arxiv.org/abs/1711.01076}{{\ttfamily arXiv:1711.01076 [hep-th]}}.

\bibitem{Hristov:2018lod}
K.~Hristov, I.~Lodato, and V.~Reys, ``{On the quantum entropy function in 4d
  gauged supergravity},'' \href{http://dx.doi.org/10.1007/JHEP07(2018)072}{{\em
  JHEP} {\bfseries 07} (2018) 072},
\href{http://arxiv.org/abs/1803.05920}{{\ttfamily arXiv:1803.05920 [hep-th]}}.

\bibitem{Liu:2018bac}
J.~T. Liu, L.~A. Pando~Zayas, and S.~Zhou, ``{Subleading Microstate Counting in
  the Dual to Massive Type IIA},''
\href{http://arxiv.org/abs/1808.10445}{{\ttfamily arXiv:1808.10445 [hep-th]}}.

\bibitem{Gang:2019uay}
D.~Gang, N.~Kim, and L.~A. Pando~Zayas, ``{Precision Microstate Counting for
  the Entropy of Wrapped M5-branes},''
  \href{http://dx.doi.org/10.1007/JHEP03(2020)164}{{\em JHEP} {\bfseries 03}
  (2020) 164}, \href{http://arxiv.org/abs/1905.01559}{{\ttfamily
  arXiv:1905.01559 [hep-th]}}.

\bibitem{PandoZayas:2019hdb}
L.~A. Pando~Zayas and Y.~Xin, ``{Topologically twisted index in the 't Hooft
  limit and the dual AdS$_4$ black hole entropy},''
  \href{http://dx.doi.org/10.1103/PhysRevD.100.126019}{{\em Phys.\ Rev.\ D}
  {\bfseries 100} no.~12, (2019) 126019},
  \href{http://arxiv.org/abs/1908.01194}{{\ttfamily arXiv:1908.01194
  [hep-th]}}.

\bibitem{Hristov:2019xku}
K.~Hristov, I.~Lodato, and V.~Reys, ``{One-loop determinants for black holes in
  4d gauged supergravity},''
  \href{http://dx.doi.org/10.1007/JHEP11(2019)105}{{\em JHEP} {\bfseries 11}
  (2019) 105}, \href{http://arxiv.org/abs/1908.05696}{{\ttfamily
  arXiv:1908.05696 [hep-th]}}.

\bibitem{GonzalezLezcano:2020yeb}
A.~Gonz\'alez~Lezcano, J.~Hong, J.~T. Liu, and L.~A. Pando~Zayas,
  ``{Sub-leading Structures in Superconformal Indices: Subdominant Saddles and
  Logarithmic Contributions},''
  \href{http://arxiv.org/abs/2007.12604}{{\ttfamily arXiv:2007.12604
  [hep-th]}}.

\bibitem{PandoZayas:2020iqr}
L.~A. Pando~Zayas and Y.~Xin, ``{Universal Logarithmic Behavior in Microstate
  Counting and the Dual One-loop Entropy of AdS$_4$ Black Holes},''
  \href{http://arxiv.org/abs/2008.03239}{{\ttfamily arXiv:2008.03239
  [hep-th]}}.

\bibitem{Ghosh:2020rwf}
J.~K. Ghosh and L.~A. Pando~Zayas, ``{Comments on Sen's Classical Entropy
  Function for Static and Rotating AdS$_4$ Black Holes},''
  \href{http://arxiv.org/abs/2009.11147}{{\ttfamily arXiv:2009.11147
  [hep-th]}}.

\bibitem{Sen:2005wa}
A.~Sen, ``{Black hole entropy function and the attractor mechanism in higher
  derivative gravity},''
  \href{http://dx.doi.org/10.1088/1126-6708/2005/09/038}{{\em JHEP} {\bfseries
  09} (2005) 038}, \href{http://arxiv.org/abs/hep-th/0506177}{{\ttfamily
  arXiv:hep-th/0506177}}.

\bibitem{Hubeny:2009ru}
V.~E. Hubeny, D.~Marolf, and M.~Rangamani, ``{Hawking radiation in large N
  strongly-coupled field theories},''
  \href{http://dx.doi.org/10.1088/0264-9381/27/9/095015}{{\em Class. Quant.
  Grav.} {\bfseries 27} (2010) 095015},
  \href{http://arxiv.org/abs/0908.2270}{{\ttfamily arXiv:0908.2270 [hep-th]}}.

\bibitem{Hubeny:2009rc}
V.~E. Hubeny, D.~Marolf, and M.~Rangamani, ``{Hawking radiation from AdS black
  holes},'' \href{http://dx.doi.org/10.1088/0264-9381/27/9/095018}{{\em Class.
  Quant. Grav.} {\bfseries 27} (2010) 095018},
  \href{http://arxiv.org/abs/0911.4144}{{\ttfamily arXiv:0911.4144 [hep-th]}}.

\bibitem{Copetti:2020dil}
C.~Copetti, A.~Grassi, Z.~Komargodski, and L.~Tizzano, ``{Delayed Deconfinement
  and the Hawking-Page Transition},''
  \href{http://arxiv.org/abs/2008.04950}{{\ttfamily arXiv:2008.04950
  [hep-th]}}.

\bibitem{Page:1993wv}
D.~N. Page, ``{Information in black hole radiation},''
  \href{http://dx.doi.org/10.1103/PhysRevLett.71.3743}{{\em Phys. Rev. Lett.}
  {\bfseries 71} (1993) 3743--3746},
  \href{http://arxiv.org/abs/hep-th/9306083}{{\ttfamily arXiv:hep-th/9306083}}.

\bibitem{Page:2013dx}
D.~N. Page, ``{Time Dependence of Hawking Radiation Entropy},''
  \href{http://dx.doi.org/10.1088/1475-7516/2013/09/028}{{\em JCAP} {\bfseries
  09} (2013) 028}, \href{http://arxiv.org/abs/1301.4995}{{\ttfamily
  arXiv:1301.4995 [hep-th]}}.

\bibitem{Nian:2019buz}
J.~Nian, ``{Kerr Black Hole Evaporation and Page Curve},''
  \href{http://arxiv.org/abs/1912.13474}{{\ttfamily arXiv:1912.13474
  [hep-th]}}.

\end{thebibliography}\endgroup
\bibliographystyle{utphys}

\end{document}